# On the Principled Description of Human Movements

**Stuart Hagler**

Oregon Health & Science University

Portland, OR, USA

haglers@ohsu.edu

**Abstract.** While the use of technology to provide accurate and objective measurements of human movement performance is presently an area of great interest, efforts to quantify the performance of movement are hampered by the lack of a principled model that describes how a subject goes about making a movement. We put forward a principled mathematical formalism that describes human movements using an optimal control model in which the subject controls the jerk of the movement. We construct the formalism by assuming that the movement a subject chooses to make is better than the alternatives. We quantify the relative quality of movements mathematically by specifying a cost functional that assigns a numerical value to every possible movement; the subject makes the movement that minimizes the cost functional. We develop the mathematical structure of movements that minimize a cost functional, and observe that this development parallels the development of analytical mechanics from the Principle of Least Action. We derive a constant of the motion for human movements that plays a role that is analogous to the role that the energy plays in classical mechanics. We apply the formalism to the description of two movements: (1) rapid, targeted movements of a computer mouse, and (2) finger-tapping, and show that the constant of the motion that we have derived provides a useful value with which we can characterize the performance of the movements. In the case of rapid, targeted movements of a computer mouse, we show how the model of human movement that we have developed can be made to agree with Fitts' law, and we show how Fitts' law is related to the constant of the motion that we have derived. We finally show that solutions exist within the model of human movements that exhibit an oscillatory character reminiscent of tremor.

**Acknowledgments:** This work was supported by the Alzheimer's Association, the National Institute of Health, the National Institute on Aging, the National Institute on Standards and Technology's Advanced Technology Projects, and the National Science Foundation under the Grants 1111722, 5RC1AG36121-2, ASMMI0116ST, P30AG008017, P30AG024978, and R01AG024059. This work was also partially funded by Intel Corporation. The content is solely the responsibility of the author and does not necessarily represent the official views of the Alzheimer's Association, the National Institute of Health, the National Institute on Aging, the National Institute on Standards and Technology's Advanced Technology Projects, the National Science Foundation, or the Intel Corporation.

# Table of Contents

# Chapter 1 – An Optimal Jerk-Control Model of Human Movements

**Abstract.** While the use of technology to provide accurate and objective measurements of human movement performance is presently an area of great interest, efforts to quantify the performance of movement are hampered by the lack of a principled model that describes how a subject goes about making a movement. We put forward a principled model of human movements using an optimal control model in which the jerk (the time derivative of the acceleration) is controlled. We use a variational approach that is a generalization of analytical mechanics (an approach to classical mechanics following from the calculus of variations). We derive Lagrangian and Hamiltonian formalisms for this model analogous to the Lagrangian and Hamiltonian formalisms in analytical mechanics. We also derive a conservation law for the model of human movements that plays a role analogous to that of conservation of energy in classical mechanics. In an effort to build intuition as to how the model of human movements works, we proceed by keeping close contact to classical mechanics and noting the parallels between the two.

## 1.1 Introduction

We can obtain much information about a person's cognitive and physical performance by observing how that person moves their body. Historically, this information has been obtained by having a trained expert, whether a clinician or an athletic coach, observe a person and provide an assessment of their performance and an enumeration of any issues with the performance that the expert may have observed. Unfortunately, the reliance on an expert introduces elements subjectivity into the performance assessment. The development of a variety of technologies has allowed the possibility of moving beyond performance assessments based on observations made by experts to more objective, and detailed measurements of the movements themselves. However, the performance of technology in providing objective measurements of motor performance has been limited by the lack of a principled model of how a person decides what movements to make and how the person carries those movements out. When measuring the performance of a movement, a principled model of movement tells one what should be measured, why it should be measured, and what the measurement means, whereas without a principled model one is left guessing at the proper measurements to make.

Measurements of human movements have been used as a means to characterize cognitive and motor performance in a variety of settings. A variety of techniques of performance assessment have been employed in sports biomechanics with the aim of improving athletic performance (see e.g. [1, 2]), and much of the growing field of wearable technology relies on the measurement of movements. [3] Measurements of human movements have also been used as the basis of performance estimates made remotely using sensors placed in the home. Some examples of in-home measurement include walking speed, [4-6] automated exercise coaching, [7] computer usage, [8] typing speed, [9] play of a computer game, [10] everyday computer mouse usage, [11] and sleep habits [12]. One commonality of these in-home measurement techniques is that they extrapolate estimates of cognitive and motor performance from measurement of motor behaviors made in the home. Nevertheless, they lack detailed models describing the motor behaviors that they measure, instead being content with summary measures like walking speed [6] or Fitts' law [10]. The lack of a detailed model of the measured motor behaviors limits the ability of these techniques to make good measurements from the range and variety of motor behaviors present in life.

A natural starting point for constructing a mathematical model of human motor behavior is to assume that movements are optimal in some sense. [13] Classical control methods are able to calculate the sequence of positions as a function of time for a movement by calculating the sequence of position that minimizes the value of some cost functional. [14-19] Unfortunately, there are no general principles for determining the form of the cost functional to be used in finding the movement a human will make a specific situation. [13] However, one can formulate the motor behavior problem (even for complicated movements) as a problem in which one calculates movements of a few key points. [20-24] In this case, one assumes that the body is constrained so that the movement of those key points determines the movement of the remainder of the body.

In the following, we construct an optimal control model for motor behavior. To keep the analysis straightforward and intuitive, we construct a model for basic one-dimensional movements such as moving the hand from one place to another. The model should extend to more complicated examples of motor behavior including quiet standing and walking. The approach is to construct a model using a variational approach that parallels the variational approach used in physics to construct classical mechanics using Lagrangian and Hamiltonian methods (see e.g., [25] Chapters 2 and 8, an accessible, popular account of classical mechanics that goes over variational methods can be found in [26]). The model assumes that a particular point on the body is controlled and that the rest of the body moves as it must to facilitate the motion of the controlled point as in [20-24]. We assume that motor behaviors are optimal in the sense of minimizing a cost functional that is analogous to the action in classical mechanics. The aspect of the movement that is specifically controlled is the jerk; [15-18] this aspect of the movement plays a role in the model that is analogous to the role the velocity plays in classical mechanics. We construct both Lagrangian and Hamiltonian formulations of the optimal control problem that determines the form of the motor behavior. We also derive a conservation law that plays a role in the model of human movement that is analogous to the role that the law of conservation of energy plays in classical mechanics.

**1.2 Motivation**

We begin with a thought-experiment. The subject is seated in a chair and holds an arm outstretched so that the hand is at some position A in space. The subject holds the hand motionless at A for some short time-interval before some trigger (e.g., a bell) indicates that they must move the hand to another position B in space. After moving the hand to position B, the subject holds the hand motionless for another short time-interval. We record the sequence of positions the hand takes in moving from position A to position B at some suitably high sampling rate; this sequence of positions moving from A to B is called the *orbit* of the movement. We perform this experiment on a cohort of normal, healthy subjects at regular intervals over some extended period providing measurements of the orbit for many subjects across an extended period.

When we conclude the series of experiments, we compare the measured orbits. From experience, we expect that the observed orbits are going to be reasonably similar. Subjects will move the in more or less a straight-line from point A to point B without wild excursions. We do not expect a subject to move the hand from point A, to the right knee, and then to the nose, and finally to point B unless they are teasing the experimenter. In addition, the subject makes movement smoothly with the velocity increasing from zero to some peak during the first part of the movement and falling back to zero toward the end. We do not expect the subject to start moving from A to B, then start moving back to A, before again moving toward B and finally completing the movement.

The question is why subjects choose to make reasonably similar orbits in moving the hand from position A to position B when they are free to choose from among a wide range of orbits that begin and end at the same positions. We argue that the orbit the subjects choose is in some measurable sense better than the alternatives. Subjects are just choosing the best of the possible orbits in the sense of moving with the orbit that is associated with the best value of the measure.

When positions A and B are a comfortable distance apart (at most maybe a foot apart) and located in space relative to the subject so that the subject can make the movement from position A to position B comfortably, we expect the chosen orbit will lie very close to the straight line connecting A and B. In these cases, the problem reduces approximately to the one-dimensional problem of finding the way the orbit changes position in time along the straight-line connecting A and B.

In the mathematical analysis that follows, we assume that we are dealing with comfortably movements that move very near to the straight line connecting the initial and final positions. The analysis deals with one-dimensional movement and aims to calculate the form of the orbit as a function of time for these movements; however, we can extend the formalism to more general kinds of movement. We assume that it is not a part of the body that is being controlled (in our example, the hand and arm), but rather a point (or, for more complicated movements, points) in the body that is being controlled (in our example, a point in the hand). We assume the rest of the body is constrained to move the way it needs to so that the point being controlled carries out the chosen orbit. Put another way, we are assuming that we can the motion of the entire body given a particular orbit of the point that is controlled. Mathematically, we solve for the motion of a single point in space. The problem of identifying what point the subject controls in a particular movement is outside the scope of the present analysis, but the reader can consult [20-24].

We call the measure used to determine the best possible orbit the *cost functional* and assume that the best orbit is the one that makes the value of the cost functional a minimum. We call the best orbit, the one that minimizes the cost functional, the *optimal orbit*.

### 1.3 Jerk-Control Cost Functional

Mathematically, we write the general cost functional for an orbit $x(t)$ moving in one-dimension as:

$$J[x] = \int_0^T L(x, dx/dt, \ldots, d^m x/dt^m) dt. \tag{1.1}$$

We assume that the movement begins at time 0 and end at time $T$. We calculate cost of the movement by taking the time-integral of some function of the orbit and its time-derivatives up to order $m$. We call this function, the integrand of Eq. (1.1), the *Lagrangian*. The orbit that minimizes the value of the integral in Eq. (1.1) is the optimal orbit, that is, the orbit by which the subject carries out the movement.

We assume there is some quantity the subject controls perfectly, and that this perfectly controlled quantity determines the orbit $x(t)$ of the movement. By "control perfectly," we mean that the subject can set the value of this controlled quantity at will, discontinuously. Intuitively, this means that the controlled quantity can behave as a knob that, when turned, passes through all the intermediate values between the initial and final position, or a switch that moves discontinuously from one discrete value to another. Moreover, we assume that the controlled quantity is in fact one of the time-derivatives of the orbit - $d^n x/dt^n$.

As the control $d^n x / dt^n$ is potentially discontinuous, this means that time-derivatives of the control potentially contain points at infinity where the time-derivative is taken across a discontinuity. We argue that this pathological mathematical behavior makes derivatives of higher order than the control not generally physically meaningful, and that we should limit the Lagrangian to derivatives of at most the order of the control:

$$J[x] = \int_0^T L(x, dx/dt, \ldots, d^n x/dt^n) dt. \tag{1.2}$$

We now need to decide the order $n$ of the controlled quantity $d^n x / dt^n$ for motor behaviors. As we have argued that $d^n x / dt^n$ can be discontinuous, the order $n$ corresponds to the lowest order of time-derivative of the orbit $x(t)$ at which we are willing to tolerate discontinuities. Physically, a human being moves a part of the body using forces exerted by muscles. We argue that the forces generated by muscles cannot discontinuous, but that when the force exerted by muscles changes from one value to another it must pass through all intervening values. Since, according to Newton's second law, a force on a body accelerates that body, this means that the accelerations generated by those muscles cannot be discontinuous. Thus, the control must be of order $n > 2$. The most straightforward (i.e. lowest order) case, then is to take $n = 3$ and use the jerk as the control. Using dots to denote the time-derivatives, the cost functional for motor behavior takes the form:

$$J[x] = \int_0^T L(x, \dot{x}, \ddot{x}, \dddot{x}) dt. \tag{1.3}$$

We simplify further by assuming there is no interaction in the cost functional between the controlled jerk and the position, velocity, or acceleration of the orbit. This allows us to treat the Lagrangian as the sum of two functions:

$$J[x] = \int_0^T \left( M(\dddot{x}) - \Phi(x, \dot{x}, \ddot{x}) \right) dt. \tag{1.4}$$

The negative sign in the integrand Eq. (1.4) appears by analogy to classical mechanics. As $\mathrm{M}(\ddot{x})$ represents a physical quantity, we expect it to be sufficiently well-behaved that we can approximate it by a truncated Taylor series expansion. Truncating $\mathrm{M}(\ddot{x})$ to second-order gives:

$$\mathrm{M}(\ddot{x}) \approx \mu_0 + \mu_1 \ddot{x} + \frac{1}{2}\mu_2 \ddot{x}^2. \tag{1.5}$$

We will not prove it here, although it is a consequence of Eq. (1.19), the first two terms on the right-hand side (RHS) of Eq. (1.5) do not affect the calculation of the optimal orbit and can be ignored. As we do not have any explicit measurement for $\mu_2$ we can suppress it by dividing through Eq. (1.4) by $\mu_2$ and redefine $\Phi(x, \dot{x}, \ddot{x})$ accordingly. The cost functional now takes the form:

$$J[x] = \int_0^T \left( \frac{1}{2}\ddot{x}^2 - \Phi(x, \dot{x}, \ddot{x}) \right) dt. \tag{1.6}$$

The process used for deriving an approximate form of $\frac{1}{2}\mu_2 \ddot{x}^2$ from $\mathrm{M}(\ddot{x})$ is analogous to the process by which the kinetic energy $\frac{1}{2}m\dot{x}^2$ of classical mechanics can be derived as an approximation to special relativity (see e.g., [25] Chapter 7, or [27]). Classical mechanics is a low velocity approximation to special relativity; and we expect Eq. (1.6) to be a low jerk approximation to a more correct Eq. (1.4).

We have argued that the accelerations during the orbit $x(t)$ correspond to the net forces exerted by the muscles. If we replace the accelerations $\ddot{x}(t)$ in the orbit with the associated forces $f(t) \sim \ddot{x}(t)$, we arrive at a cost functional associated with the cost of muscle activations.

For convenience of expression, we use the notation of $f^{(n)}$ to denote the $n$-th integral of $f$ with respect to time. Using this notation, we may reformulate Eq. (1.6) as:

$$J[f] = \int_0^T \left( \frac{1}{2} \dot{f}^2 - \Phi\left(f^{(2)}, f^{(1)}, f\right) \right) dt. \tag{1.7}$$

Eq. (1.7) provides a formulation of the problem in the force domain. It has a closer resemblance to the cost functional used in classical mechanics as the first term in the integrand involves a square of the first time-derivative of a quantity (here the net force exerted by the muscles). That is, when looked at in the force domain, the model looks formally more like classical mechanics (see e.g., [25] Chapter 2).

**1.4 Alternative Control Models**

Although we have argued for jerk-control in Section 1.3, we can develop intuition for what jerk control is doing by looking at alternatives for the quantity $d^n x / dt^n$ that is controlled in a movement. We can do this by repeating the derivation laid out in Section 1.3 for any choice of control order $n$. To keep the analysis straightforward, we take the function $\Phi\left(x, dx/dt, \ldots, d^{n-1}x/dt^{n-1}\right)$ to be zero in all cases.

The general formulation of the optimal control problem in this case is to find the orbit $x(t)$ that minimizes the cost functional $J[\cdot]$ subject to the indicated initial and final conditions:

$$\begin{aligned}
J[x] &= \int_0^1 \frac{1}{2} \left( \frac{d^n x}{dt^n} \right)^2 dt, \\
x(0) &= -1, \quad x(1) = 0, \\
&\ldots, \quad \ldots, \\
\frac{d^{n-1} x}{dt^{n-1}}(0) &= 0, \quad \frac{d^{n-1} x}{dt^{n-1}}(1) = 0.
\end{aligned} \tag{1.8}$$

The movement travels from an initial position -1 to a final position at the origin during the time interval $0 \leq t \leq 1$. We suppose the movement, as in Section 1.2, to be motionless for some small time interval at the beginning and end of the movement, so all the time derivatives of the orbit $x(t)$ are zero at the beginning and end of the movement. Although we do not prove it here, it follows from Eq. (1.19); the optimal orbit for the cost functional in Eq. (1.8) takes the form:

$$x(t) = \sum_{m=0}^{2n} c_m t^m. \qquad (1.9)$$

We calculate the constants $c_m$ by imposing the initial and final conditions in Eq. (1.8) on Eq. (1.9).

In Figure 1.1, we give the curves of the orbit $x(t)$, its velocity $\dot{x}(t)$, and its acceleration $\ddot{x}(t)$ for Eq. (1.9) using orders $n$ of 2 (acceleration-control), 3 (jerk-control), 4 (snap-control), 5 (crackle-control), and 6 (pop-control). The movement travels from position -1 to the origin the time interval $0 \leq t \leq 1$; it is motionless for short intervals before and after the movement. Figure 1.1 illustrates how the optimal orbit changes its form as the order of control is changed to higher order derivatives of the orbit. The reader should note the curve for the acceleration of the orbit $\ddot{x}(t)$ when the control order $n$ is 2 (acceleration-control). There are discontinuities in the acceleration of the orbit at times 0 and 1. This corresponds to the control being able to behave as a switch as discussed in Section 1.3. We have chosen to build a model in which jerk is controlled in order to avoid this kind of behavior in the orbit $x(t)$, its velocity $\dot{x}(t)$, or its acceleration $\ddot{x}(t)$.

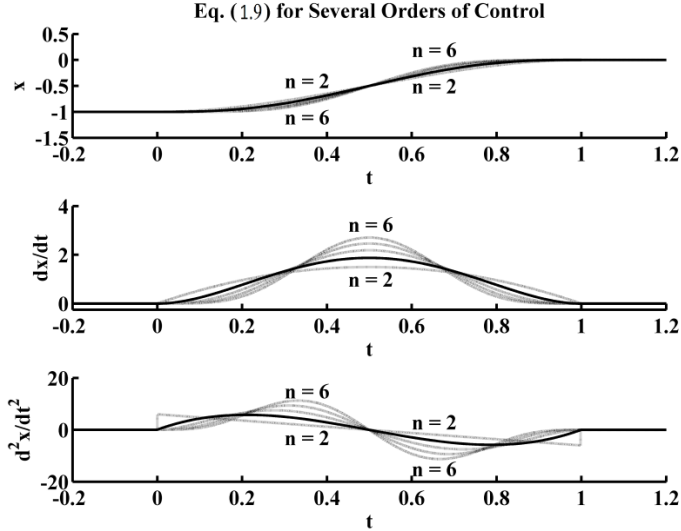

**Figure 1.1.** Eq. (1.9) for several orders of control. We compare the curves of the orbit $x(t)$, its velocity $\dot{x}(t)$, and its acceleration $\ddot{x}(t)$ for Eq. (1.9) using orders $n$ of 2 (acceleration-control), 3 (jerk-control), 4 (snap-control), 5 (crackle-control), and 6 (pop-control). The movement travels from position -1 to the origin the time interval $0 \leq t \leq 1$; it is motionless for short intervals before and after the movement. The curves for $n$ of 2 and 6 are indicated, the curves regularly move from $n$ of 2 to $n$ of 6 as the order is increased. The curve for $n$ of 3 (jerk-control) is indicated in bold.

## 1.5 Lagrangian Mechanics of Optimal Jerk-Control

We now develop the mechanics of the Lagrangian formulation of the optimal jerk-control movement problem. In the Lagrangian mechanics formalism, the optimal solution to the movement problem is given by a single sixth-order differential equation – the Euler-Lagrange equation. After deriving the formula of the optimal solution, we proceed to analyze the formulation of the solution into quantities that generalize established quantities from classical mechanics, namely, force, momentum, energy, kinetic energy, and potential energy. We develop these quantities anticipating that the generalized versions of these quantities may prove as useful in the optimal jerk-control movement mechanics as they have proven to be in classical mechanics. In addition, we show that the solution to the Euler-Lagrange equation has an associated conservation law that generalizes the law of conservation of energy in classical mechanics.

Eq. (1.6) gives the cost functional for jerk-controlled movements. We now set out the mathematics of calculating the optimal orbit $x(t)$ for a cost functional of that form. This mathematical formalism includes calculating form of the optimal orbit itself as well as working out some useful mathematical properties that the optimal orbit will necessarily possess. What follows is an elaboration of the mathematical structure of optimizing a cost functional; we do not introduce any new physical assumptions.

As we indicated in Eq. (1.8) in without further elaboration, we specify a movement by indicating the initial and final conditions that it must satisfy. The optimal movement is the one that starts in the initial conditions, ends in the final conditions, and minimizes the cost functional. Mathematically, it happens that for control of order 3 (i.e. jerk-control), we must specify six initial and final conditions. We assume a movement starts at some position $-D$ and concludes at the origin in a time interval $0 \leq t \leq T$, this fixes two of the initial and final conditions leaving. The easiest way to specify the remaining four initial and final conditions is to specify the initial and final velocities and accelerations. Following this, we give the general form of the optimization problem for jerk-control of a *simple movement* as one of minimizing the cost functional $J[\cdot]$ while having the specified initial and final conditions:

$$J[x] = \int_0^T \left( \frac{1}{2}\dddot{x}^2 - \Phi(x, \dot{x}, \ddot{x}) \right) dt,$$
$$x(0) = -D, \quad x(T) = 0,$$
$$\dot{x}(0) = v_0, \quad \dot{x}(T) = v_1, \qquad (1.10)$$
$$\ddot{x}(0) = a_0, \quad \ddot{x}(T) = a_1.$$

We define a *basic movement* to be a simple movement that begins and ends at rest. Thus, for basic movements, the initial and final velocities and accelerations are zero. Setting the initial and final velocities to zero means that the point that the subject controls is not moving at the beginning and ending of the movement, while setting the initial and final accelerations to zero means that the point being controlled is not moving shortly before the beginning of the movement or shortly after the end of the movement.

The formalism that we are developing for jerk-controlled movement is just a generalization of the formalism for classical mechanics. To motivate intuition particular cases, we at times look at the analogous cases in classical mechanics. The classical mechanics problem we look at is the motion of a point particle in an external field (e.g., gravity) in one dimension; the problem is to find the orbit $x(t)$ that makes the cost functional $S[\cdot]$, the *action*, stationary (i.e. having a first variation of zero) while having certain specified initial and final positions:

$$S[x] = \int_0^T L(x, \dot{x}) dt = \int_0^T \left( \frac{1}{2} \dot{x}^2 - \Phi(x) \right) dt, \qquad (1.11)$$
$$x(0) = -D, \quad x(T) = 0.$$

The supposition that a physical movement makes the action in Eq. (1.11) stationary is known as the *Principle of Least Action*. In Eq. (1.11), we have suppressed the particle's mass $m$ by dividing through by $m$ in a manner analogous to the way $\mu_2$ was suppressed in producing Eq. (1.6). The function $\Phi(x)$ describes how an external field that is affecting the motion of the particle varies throughout space.

*1.5.1 Euler-Lagrange Equation*

We now calculate the mathematical form of the optimal orbit (i.e. the orbit that minimizes the cost functional) $x^*(t)$ given a cost functional $J[\cdot]$ that is assumed to have a minimum. We denote the optimal orbit by $x^*(t)$ rather than by $x(t)$ in this section solely for mathematical clarity. We will return to using $x(t)$ for the optimal orbits in the remaining sections.

We suppose we are making a movement that must minimize the value of a jerk-control cost functional while satisfying some initial and final conditions as in Eq. (1.10). The cost $J^*$ associated with the optimal orbit $x^*(t)$ is:

$$J^* = \int_0^T L(x^*, \dot{x}^*, \ddot{x}^*, \dddot{x}^*) dt. \tag{1.12}$$

The optimal orbit $x^*(t)$ satisfies the initial and final conditions by definition. There is a family of orbits $x(t)$ near to the optimal orbit and satisfying initial and final conditions. We may express these orbits mathematically using a function $\delta x(t)$ so that:

$$x(t) = x^*(t) + \delta x(t). \tag{1.13}$$

We express the time derivatives of the function $\delta x(t)$ using the notation:

$$\delta \dot{x} = \frac{d\delta x}{dt}, \quad \delta \ddot{x} = \frac{d\delta \dot{x}}{dt}, \quad \delta \dddot{x} = \frac{d\delta \ddot{x}}{dt}. \tag{1.14}$$

Since $x^*(t)$ already satisfies the initial and final conditions, and we require all orbits $x(t)$ to satisfy the initial and final conditions, the $\delta x(t)$ must satisfy:

$$\begin{aligned} \delta x(0) &= 0, & \delta x(T) &= 0, \\ \delta \dot{x}(0) &= 0, & \delta \dot{x}(T) &= 0, \\ \delta \ddot{x}(0) &= 0, & \delta \ddot{x}(T) &= 0. \end{aligned} \tag{1.15}$$

In Figure 1.2, we illustrate the relationship between the optimal orbit $x^*(t)$ and one possible orbit $x(t)$ for a movement from position -1 to the origin during a time interval $0 \leq t \leq 1$. The reader should note the way the orbits converge at $t = 0$ and $t = 1$, and imagine similar plots for other choices of $x(t)$. Using all the possible functions $\delta x(t)$, we can arrive at all the possible orbits satisfying the initial and final conditions.

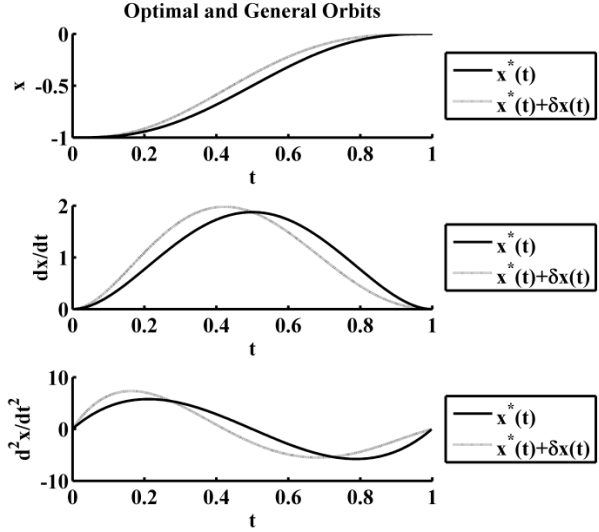

**Figure 1.2.** Optimal and general orbits. We compare the curves of the orbit $x(t)$, its velocity $\dot{x}(t)$, and its acceleration $\ddot{x}(t)$ for the optimal jerk-control orbit (solid line) and an alternative general orbit (dotted line). The movement travels from position -1 to the origin the time interval $0 \leq t \leq 1$; it is motionless for short intervals before and after the movement. The reader can imagine further general orbits that complete the desired movement.

We next calculate the first variation of the cost functional $J[\cdot]$. We do this by expanding the Lagrangian in a Taylor series around the optimal orbit $x^*(t)$:

$$J[x] = J^* + \int_0^T \left( \frac{\partial L}{\partial x} \delta x + \frac{\partial L}{\partial \dot{x}} \delta \dot{x} + \frac{\partial L}{\partial \ddot{x}} \delta \ddot{x} + \frac{\partial L}{\partial \dddot{x}} \delta \dddot{x} \right) dt + \ldots \quad (1.16)$$

The second term on the RHS of Eq. (1.16) is the first variation; further terms in the series expansion would lead the second and higher variations of the cost functional. Although the notation in Eq. (1.16) does not explicitly indicate it, we calculate the partial derivatives of the Lagrangian using the optimal orbit $x^*(t)$.

The optimal orbit $x^*(t)$ minimizes the cost functional. At the minimum, the first variation must be zero:

$$\int_0^T \left( \frac{\partial L}{\partial x} \delta x + \frac{\partial L}{\partial \dot{x}} \delta \dot{x} + \frac{\partial L}{\partial \ddot{x}} \delta \ddot{x} + \frac{\partial L}{\partial \dddot{x}} \delta \dddot{x} \right) dt = 0. \tag{1.17}$$

We may integrate Eq. (1.17) using repeated application of integration by parts, giving the equation:

$$\left( \frac{\partial L}{\partial \dddot{x}} \delta \ddot{x} \right)\Big|_0^T + \left( \left( \frac{\partial L}{\partial \ddot{x}} - \frac{d}{dt} \frac{\partial L}{\partial \dddot{x}} \right) \delta \dot{x} \right)\Big|_0^T + \left( \left( \frac{\partial L}{\partial \dot{x}} - \frac{d}{dt} \frac{\partial L}{\partial \ddot{x}} + \frac{d^2}{dt^2} \frac{\partial L}{\partial \dddot{x}} \right) \delta x \right)\Big|_0^T$$
$$+ \int_0^T \left( \frac{\partial L}{\partial x} - \frac{d}{dt} \frac{\partial L}{\partial \dot{x}} + \frac{d^2}{dt^2} \frac{\partial L}{\partial \ddot{x}} - \frac{d^3}{dt^3} \frac{\partial L}{\partial \dddot{x}} \right) \delta x \, dt = 0. \tag{1.18}$$

Using Eq. (1.15), we find that the first three terms on the left-hand side (LHS) in Eq. (1.18) are zero. This means that the fourth term on the LHS must also be zero. Since Eq. (1.18) must hold for arbitrary $\delta x$, the integrand must be zero, that is:

$$\frac{d^3}{dt^3} \frac{\partial L}{\partial \dddot{x}} - \frac{d^2}{dt^2} \frac{\partial L}{\partial \ddot{x}} + \frac{d}{dt} \frac{\partial L}{\partial \dot{x}} - \frac{\partial L}{\partial x} = 0. \tag{1.19}$$

Thus, the optimal orbit $x^*(t)$ Eq. (1.19) must satisfy Eq. (1.19). This equation is the Euler-Lagrange equation for the cost functional $J[\cdot]$. We note that Eq. (1.19) depends only on the Lagrangian. A mathematically rigorous treatment of the calculus of variations that provides a derivation of Eq. (1.19) is provided in [28]. For comparison, the Euler-Lagrange equation for classical mechanics is (see e.g., [25] Chapter 2):

$$\frac{d}{dt} \frac{\partial L}{\partial \dot{x}} - \frac{\partial L}{\partial x} = 0. \tag{1.20}$$

*1.5.2 Generalized Force & Momentum*

In classical mechanics, two concepts that have proven useful for helping people understand how a system works are: (1) the classical momentum, and (2) the classical force. We develop here the concepts of the generalized momentum and the generalized force in the expectation that they may prove as helpful in understanding how the jerk-control system works as the classical momentum and the classical force have proven in understanding how classical mechanics works.

We look first at how the classical momentum and the classical force fall out of the Euler-Lagrange equation for classical mechanics. We can calculate Newton's second law for the classical mechanics motion of a point-particle in an external field by finding the Euler-Lagrange equation for the classical mechanics Lagrangian for a point-particle in an external field (Eq. (1.11)). This gives:

$$\ddot{x} = -\frac{\partial \Phi}{\partial x}. \tag{1.21}$$

Recalling that we have suppressed the particle's mass $m$, we observe that Eq. (1.21) gives a relationship between an acceleration of the particle and the local rate of change of the value of the external field with position. The RHS of Eq. (1.21) gives the classical force $f$ of the external field on the particle as a function of the position of the particle in space. Defining the (mass-suppressed) classical momentum to be $p = \dot{x}$, we can rewrite Eq. (1.21) in the more typical form for Newton's second law as:

$$\dot{p} = f. \tag{1.22}$$

Using the derivation of the classical force and momentum from the classical mechanics Lagrangian as a guide, we now proceed to derive the generalized momentum and force from the jerk-control Lagrangian. Applying the Euler-Lagrange equation in Eq. (1.19) to the cost functional for jerk-control in Eq. (1.6) gives:

$$\dddot{x} = -\frac{\partial \Phi}{\partial x} + \frac{d}{dt}\frac{\partial \Phi}{\partial \dot{x}} - \frac{d^2}{dt^2}\frac{\partial \Phi}{\partial \ddot{x}}. \tag{1.23}$$

We can rewrite Eq. (1.23) in the form:

$$\frac{d}{dt}\left(\dddot{x} - \frac{\partial \Phi}{\partial \dot{x}} + \frac{d}{dt}\frac{\partial \Phi}{\partial \ddot{x}}\right) = -\frac{\partial \Phi}{\partial x}. \tag{1.24}$$

Eq. (1.24) relates the rate of change of the quantity on the LHS to the quantity on the RHS. Comparison Eq. (1.24) to Newton's second law in Eq. (1.22) suggests that LHS of Eq. (1.24) corresponds to the rate of change of some "momentum" while the RHS corresponds to some "force." We write these quantities as:

$$p_1 = \dddot{x} - \frac{\partial \Phi}{\partial \dot{x}} + \frac{d}{dt}\frac{\partial \Phi}{\partial \ddot{x}}, \quad f_1 = -\frac{\partial \Phi}{\partial x}. \tag{1.25}$$

The "force" $f_1$ in Eq. (1.25) is given by the rate at which the value of the Lagrangian changes when the position varied by a very small amount and all other things are kept constant. When the "force" $f_1$ is negligible (i.e. the Lagrangian is approximately independent of the position), we can rewrite Eq. (1.25) as:

$$\frac{d}{dt}\left(-\dddot{x} - \frac{\partial \Phi}{\partial \ddot{x}}\right) \approx -\frac{\partial \Phi}{\partial \dot{x}} - p_1. \tag{1.26}$$

Again, comparison to Newton's second law in Eq. (1.22) suggests that LHS of Eq. (1.26) corresponds to the rate of change of some "momentum" while the RHS corresponds to some "force." We write these quantities as:

$$p_2 = -\dddot{x} - \frac{\partial \Phi}{\partial \ddot{x}}, \quad f_2 = \frac{\partial \Phi}{\partial \dot{x}} - p_1. \tag{1.27}$$

The "force" $f_2$ in Eq. (1.27) is the sum of two quantities. The first is rate at which the value of the Lagrangian changes when the velocity varied by a very small amount and all other things are kept constant. The second quantity is the value of the "momentum" $p_1$ which appears in Eq. (1.26) as the "force" acting on the "momentum" $p_2$. When the "forces" $f_1$ and $f_2$ are negligible (i.e. the Lagrangian is approximately independent of the position and the velocity), we can rewrite Eq. (1.27) as:

$$\dddot{x} \approx -\frac{\partial \Phi}{\partial \ddot{x}} - p_2. \qquad (1.28)$$

Once more, comparison to Newton's second law in Eq. (1.22) suggests that LHS of Eq. (1.26) corresponds to the rate of change of some "momentum" while the RHS corresponds to some "force." We write these quantities as:

$$p_3 = \ddot{x}, \quad f_3 = -\frac{\partial \Phi}{\partial \ddot{x}} - p_2. \qquad (1.29)$$

The "force" $f_3$ in Eq. (1.29) is again the sum of two quantities. The first is the rate at which the value of the Lagrangian changes when we vary the acceleration by a very small amount keeping all other things constant. The second quantity is the value of the "momentum" $p_2$ which appears in Eq. (1.29) as the "force" acting on the "momentum" $p_3$.

Defining the useful vector $X^T = \begin{bmatrix} x & \dot{x} & \ddot{x} \end{bmatrix}$, we may combine the three "momenta" just defined into a single vector – the generalized momentum:

$$P = \begin{bmatrix} p_1 \\ p_2 \\ p_3 \end{bmatrix} = \begin{bmatrix} \dddot{x} \\ -\ddot{x} \\ \ddot{x} \end{bmatrix} - \begin{bmatrix} 0 & 1 & -d/dt \\ 0 & 0 & 1 \\ 0 & 0 & 0 \end{bmatrix} \frac{\partial \Phi}{\partial X}. \qquad (1.30)$$

Similarly, we may combine the three "forces" just defined into a single generalized force taking the form of a vector – the generalized force:

$$F = \begin{bmatrix} f_1 \\ f_2 \\ f_3 \end{bmatrix} = -\frac{\partial \Phi}{\partial X} - \begin{bmatrix} 0 & 0 & 0 \\ 1 & 0 & 0 \\ 0 & 1 & 0 \end{bmatrix} P. \qquad (1.31)$$

The generalized form of Newton's second law relates the generalized momentum and force:

$$\dot{P} = F. \qquad (1.32)$$

By plugging Eqs. (1.30) and (1.31) into Eq. (1.32), we regain the Euler-Lagrange equation (Eq. (1.23)).

*1.5.3 Generalized Energy*

In addition generalizing the classical momentum and force to a generalized momentum and force, we may also generalize classical energy to a generalized energy. What makes the classical energy useful in classical mechanics is the law of conservation of classical energy – the total classical energy at any instant during a classical orbit remains constant through the motion. The classical energy provides a summary quantity that characterizes an entire orbit using a single number that allows one to compare one (entire) orbit to another. We show here that there is a quantity in the jerk-control system that is analogous to the classical energy. This quantity, the generalized energy, is constant over the course of an orbit that minimizes the jerk-control cost functional. It provides the same sort of information about the jerk-control system that the classical energy provides for classical mechanics.

There is one important difference between and orbit in classical mechanics and a human movement described a jerk-control system. In principle, classical mechanics describes the orbit for all time, so the classical energy associated with the orbit is conserved for all time. However, the orbit of the jerk-control system only describes the movement for the duration of time during which the movement takes place. This is the case even though, mathematically speaking, the orbit of the jerk-control system is defined for all time just as in the classical mechanics system.

In effect, we only use a segment of the jerk-control orbit to describe an actual human movement. As a result, the generalized energy is only conserved for the duration of the movement, and conservation fails at the very beginning and ending of the movement.

As we did in Section 1.5.2, we begin by looking at how the classical energy appears in classical mechanics. For the classical mechanics system described by the cost functional in Eq. (1.11), the total energy $E$ for the particle of mass $m$ moving in an external field satisfies:

$$\frac{1}{2}\dot{x}^2 + \Phi(x) = \frac{E}{m}. \tag{1.33}$$

Thus, the velocity of the orbit $\dot{x}(t)$ takes values based on the value of the external field at the location $x(t)$ that keeps the total energy constant. Eq. (1.33) just expresses the law of conservation of classical energy in classical mechanics. Using the classical momentum as defined in Section 1.5.2, we may rewrite the law of conservation of classical energy (Eq. (1.33)) using the Lagrangian; this gives:

$$p\dot{x} - L = \frac{E}{m}. \tag{1.34}$$

It is important to note that the law of conservation of classical energy expressed in Eqs. (1.33) and (1.34) only holds when the motion described by $x(t)$ optimizes the classical mechanics action.

Returning to the jerk-control model of human movement, we may use the properties of the Lagrangian when calculated along an optimal orbit to derive an equation that is the analog to Eq. (1.34) (see Section 1.10):

$$P^T \dot{X} - L = \Psi. \tag{1.35}$$

The constant of the motion $\Psi$ is the generalized energy, and Eq. (1.35) provides the expression for a law of conservation of generalized energy. As with the law of conservation of classical energy, the law of conservation of general energy only holds for orbits $x(t)$ that optimize the jerk-control cost functional. Writing out the LHS of Eq. (1.35), we arrive at an expression for the law of conservation of generalized energy that is analogous to the law of conservation of classical energy expressed in Eq. (1.33):

$$\frac{1}{2}\dddot{x}^2 - \dddot{x}\cdot\ddot{x} + \dddot{x}\cdot\dot{x} + \Phi - \left(\frac{\partial\Phi}{\partial\dot{x}} - \frac{d}{dt}\frac{\partial\Phi}{\partial\ddot{x}}\right)\dot{x} - \frac{\partial\Phi}{\partial\ddot{x}}\ddot{x} = \Psi. \tag{1.36}$$

*1.5.4 Kinetic & Potential Generalized Energy*

In classical mechanics it has also be found useful to divide the total classical energy into two components: (1) the kinetic classical energy, and (2) the potential classical energy. For the classical mechanics case of a particle moving in an external field, we can think of the kinetic classical energy as the energy we can measure directly from the observed orbit $x(t)$ of the particle. In the case of the classical mechanical particle, we do not observe the potential classical energy directly, but we infer it from the law of conservation of classical energy. We find ourselves in a similar situation when we analyze human movements. We can measure the orbits $x(t)$ of various points on the body, but have a more difficult time measuring what else the body is doing to create those orbits. As a result, we find it useful to divide the total generalized energy into two components in analogy to classical mechanics: (1) the kinetic generalized energy, and (2) the potential generalized energy.

We again begin by looking at the classical mechanics case of the particle moving in an external field. When divided into components of kinetic and potential classical energy, the equation for total energy for the classical mechanics particle moving in an external field in Eq. (1.33)

becomes the sum of the (mass-suppressed) kinetic classical energy $E_K/m$ and the (mass-suppressed) potential classical energy $E_P/m$ where:

$$\frac{E_K}{m} = \frac{1}{2}\dot{x}^2, \quad \frac{E_P}{m} = \Phi. \tag{1.37}$$

We reiterate that the classical kinetic energy $E_K/m$ contains the terms in Eq. (1.33) that an observer of the system can measure from the physical motion of the particle, while the classical potential energy $E_P/m$ contains that part of the classical energy that an observer of the system must infer.

We can illustrate the way we can use the kinetic classical energy to gain information about a system with unknown $\Phi(x)$ using the orbit of a planet around the Sun. We imagine in this illustration that we do not know the form of $\Phi(x)$ provided by the theory of gravity.

A planet orbits the Sun along an ellipse with the Sun at one focus as illustrated in Figure 1.3. The velocity of the planet is larger when it is nearer to the Sun and smaller when farther away from the Sun. So the classical kinetic energy as given in Eq. (1.37) becomes larger when the planet is nearer to the Sun and smaller when it is farther away. By imposing the principle of conservation of energy, we can calculate an empirical model for the classical potential energy that describes the change in classical kinetic energy over the course of the orbit.

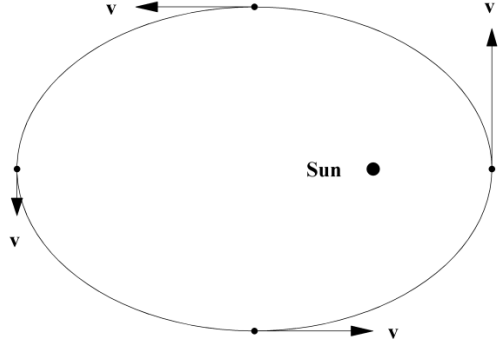

**Orbit of a Planet Around the Sun**

**Figure 1.3.** Orbit of a planet around the Sun. A planet orbits the Sun along an ellipse with changing velocity. The velocity has the highest magnitude when the planet is closest to the Sun and lowest when the planet is furthest from the Sun. We can use the velocity to calculate the kinetic classical energy of the planet during the orbit, and, by assumng conservation of classical energy, use the kinetic classical energy to characterize empirically the potential classical energy associated with the Sun's gravity.

Returning the jerk-control system, we define the kinetic generalized energy $\Psi_K$ to include all the terms in Eq. (1.36) that an observer can measure from the physical motion of the body, and we define the potential generalized energy $\Psi_P$ to be the remaining terms in Eq. (1.36). Using these definitions, the kinetic and potential generalized energies are:

$$\Psi_K = \frac{1}{2}\ddot{x}^2 - \dddot{x}\cdot\dot{x} + \dddot{x}\cdot x,$$
$$\Psi_P = \Phi - \left(\frac{\partial \Phi}{\partial \dot{x}} - \frac{d}{dt}\frac{\partial \Phi}{\partial \ddot{x}}\right)\dot{x} - \frac{\partial \Phi}{\partial \ddot{x}}\ddot{x}. \tag{1.38}$$

We may use Eqs. (1.36) and (1.38) to write an expression for the unknown $\Phi(x,\dot{x},\ddot{x})$ in terms of the known kinetic generalized energy $\Psi_K$ of:

$$(\Phi - \Psi) - \left(\frac{\partial \Phi}{\partial \dot{x}} - \frac{d}{dt}\frac{\partial \Phi}{\partial \ddot{x}}\right)\dot{x} - \frac{\partial \Phi}{\partial \ddot{x}}\ddot{x} = \Psi_K. \tag{1.39}$$

Defining $\Phi'(x,\dot{x},\ddot{x}) = \Phi(x,\dot{x},\ddot{x}) - \Psi$, that is, shifting the $\Phi(x,\dot{x},\ddot{x})$ by a constant offset equal to the generalized energy, we find:

$$\Phi' - \left(\frac{\partial \Phi'}{\partial \dot{x}} - \frac{d}{dt}\frac{\partial \Phi'}{\partial \ddot{x}}\right)\dot{x} - \frac{\partial \Phi'}{\partial \ddot{x}}\ddot{x} = \Psi_K. \tag{1.40}$$

This equation indicates what information we can obtain about $\Phi'(x,\dot{x},\ddot{x})$ given an observed movement with kinetic generalized energy $\Psi_K$ under the assumption that the jerk-control model of human movement holds.

### 1.6 Hamiltonian Mechanics of Optimal Jerk-Control

We now develop the mechanics of the Hamiltonian formulation of the optimal jerk-control movement problem. In the Hamiltonian mechanics formalism, the optimal solution to the movement problem is given by a system of six first-order differential equations – Hamilton's equations. Hamiltonian mechanics formulates the problem of making a movement in one dimension within a six dimensional phase-space consisting of three coordinate values and three momentum values. The Hamiltonian mechanics formalism is a mathematically equivalent formulation that is dual to the Lagrangian mechanics formalism; solving the problem in either formalism yields the same solution. We present both formalisms as one may better approach some problems using the Hamiltonian formalism than using the Lagrangian formalism.

The problem we are solving is to find the optimal orbit $x(t)$ that minimizes the cost functional $J[\cdot]$ while satisfying the indicated initial and final conditions:

$$\begin{aligned} J[x] &= \int_0^T \left(\frac{1}{2}\dddot{x}^2 - \Phi(x,\dot{x},\ddot{x})\right)dt, \\ x(0) &= -D, \quad x(T) = 0, \\ \dot{x}(0) &= v_0, \quad \dot{x}(T) = v_1, \\ \ddot{x}(0) &= a_0, \quad \ddot{x}(T) = a_1. \end{aligned} \tag{1.41}$$

To express Eq. (1.41) in the Hamiltonian formalism, we must rewrite it in a phase-space form. We begin this process by renaming the control $u(t) = \dddot{x}(t)$.

We next define the generalized coordinates by the vector:

$$Q^{\mathrm{T}} = \begin{bmatrix} q_1 & q_2 & q_3 \end{bmatrix}. \tag{1.42}$$

We require the generalized coordinates to satisfy the following relationship:

$$\dot{Q} = \begin{bmatrix} 0 & 1 & 0 \\ 0 & 0 & 1 \\ 0 & 0 & 0 \end{bmatrix} Q + \begin{bmatrix} 0 \\ 0 \\ u \end{bmatrix}. \tag{1.43}$$

Eqs. (1.42) and (1.43) amount to a somewhat involved way of writing $Q^{\mathrm{T}} = X^{\mathrm{T}}$ (i.e. $Q^{\mathrm{T}} = \begin{bmatrix} x & \dot{x} & \ddot{x} \end{bmatrix}$), so the three dimensions of the phase-space corresponding to the three generalized coordinates are the position, the velocity, and the acceleration. We note that the initial and final conditions in Eq. (1.41) correspond to specifying the initial and final values of the generalized coordinates, $Q(0)$ and $Q(T)$, respectively.

We finally define the generalized momentum by the vector:

$$P^{\mathrm{T}} = \begin{bmatrix} p_1 & p_2 & p_3 \end{bmatrix}. \tag{1.44}$$

The three dimensions of the portion of the phase-space corresponding to the generalized momentum are the elements of the generalized momentum vector in Eq. (1.44). The generalized momentum defined in Eq. (1.44) will turn out to be the same as the generalized momentum given in Eq. (1.30), but for now, Eq. (1.44) just gives a vector with three unknown elements.

We can now write the Lagrangian for the jerk-control system (Eq. (1.41)) in the form:

$$L = \frac{1}{2} u^2 - \Phi(Q). \tag{1.45}$$

We calculate the Hamiltonian by taking the Legendre transform (see e.g., [29]) of Eq. (1.45). The Legendre transform of the Lagrangian has the form:

$$H = P^T \dot{Q} - L = P^T \dot{Q} - \frac{1}{2}u^2 + \Phi. \qquad (1.46)$$

We note that Eq. (1.46) bears a close resemblance to Eq. (1.35); the difference is that the Hamiltonian $H$ is a function of the orbit and takes on values for any orbit, even those that are non-optimal, while the generalized energy $\Psi$ is a constant of the motion that exists only for optimal orbits. Thus, we expect that, for optimal orbits, the Hamiltonian takes on a constant value for the entire orbit.

We may now use the Hamiltonian to calculate the optimal orbit. Although we do not work it out here (see e.g., [30] Chapter 5), the optimal orbit satisfies the equations:

$$\dot{Q} = \frac{\partial H}{\partial P}, \quad \dot{P} = -\frac{\partial H}{\partial Q}, \quad \frac{\partial H}{\partial u} = 0. \qquad (1.47)$$

We call the first two equations in Eq. (1.47) *Hamilton's equations*; together they provide the expected six first-order differential equations that describe the optimal orbit that are the counterparts in the Hamiltonian formalism to the sixth-order Euler-Lagrange equation (Eq. (1.19)) in the Lagrangian formalism. It is important to stress that Eq. (1.47) has found the optimal orbit and that the remaining mathematical manipulations are just a matter of working out what the optimal solution looks like.

We now use the third equation in Eq. (1.47) to simplify the Hamiltonian by eliminating explicit reference to the control $u$ in Eq. (1.46). To do this, we first rewrite the Hamiltonian in a more convenient form:

$$H = -\frac{1}{2}(u - p_3)^2 + \left(\frac{1}{2}p_3^2 + p_1 \dot{q}_1 + p_2 \dot{q}_2 + \Phi\right). \qquad (1.48)$$

Applying the third equation in Eq. (1.47) to Eq. (1.48) gives the relationship:

$$p_3 = u. \tag{1.49}$$

Putting Eq. (1.49) into the expression of the Hamiltonian given in Eq. (1.46) allows us to express the Hamiltonian entirely in terms of the generalized coordinates and generalized momentum:

$$H = \frac{1}{2} P^{\mathrm{T}} \begin{bmatrix} 0 & 0 & 0 \\ 0 & 0 & 0 \\ 0 & 0 & 1 \end{bmatrix} P + P^{\mathrm{T}} \begin{bmatrix} 0 & 1 & 0 \\ 0 & 0 & 1 \\ 0 & 0 & 0 \end{bmatrix} Q + \Phi. \tag{1.50}$$

Using Eq. (1.49), we can rewrite Eq. (1.43) as:

$$\dot{Q} = \begin{bmatrix} 0 & 1 & 0 \\ 0 & 0 & 1 \\ 0 & 0 & 0 \end{bmatrix} Q + \begin{bmatrix} 0 & 0 & 0 \\ 0 & 0 & 0 \\ 0 & 0 & 1 \end{bmatrix} P. \tag{1.51}$$

Applying the second equation in Eq. (1.47) to Eq. (1.50), we find:

$$\dot{P} = -\frac{\partial \Phi}{\partial Q} - \begin{bmatrix} 0 & 0 & 0 \\ 1 & 0 & 0 \\ 0 & 1 & 0 \end{bmatrix} P. \tag{1.52}$$

We have stated that the Hamiltonian formulation of the problem provides a phase-space description of the motion. The state of the system at any time $t$ is given by the generalized coordinates $Q(t)$ and generalized momentum $P(t)$. For the system we are describing, that of moving in one dimension of physical space, the phase-space representation has six dimensions. The state evolves in time according to Eqs. (1.51) and (1.52). We can look at a movement as beginning in a state $Q(0)$ and $P(0)$, and evolving deterministically in time. In this case, the problem of making the movement end in the desired final conditions $Q(T)$ becomes a problem of choosing the initial value of the generalized momentum $P(0)$ so that the evolution of the system takes it to the desired final conditions $Q(T)$.

We can now calculate the form of the generalized momentum $P$. We already know from Eq. (1.49) that it has the form $P^{\text{T}} = \begin{bmatrix} p_1 & p_2 & \dot{q}_3 \end{bmatrix}$. Taking the time-derivative of Eq. (1.51) and combining it with Eq. (1.52), we find:

$$\ddot{Q} = \begin{bmatrix} 0 & 1 & 0 \\ 0 & 0 & 1 \\ 0 & 0 & 0 \end{bmatrix} \dot{Q} - \begin{bmatrix} 0 & 0 & 0 \\ 0 & 0 & 0 \\ 0 & 0 & 1 \end{bmatrix} \frac{\partial \Phi}{\partial Q} - \begin{bmatrix} 0 & 0 & 0 \\ 0 & 0 & 0 \\ 0 & 1 & 0 \end{bmatrix} P. \tag{1.53}$$

From the third row of Eq. (1.53), we can read off the value $p_2 = -\ddot{q}_3 - \partial \Phi / \partial \dot{x}$. Taking the time-derivative of Eq. (1.53) and combining it with Eq. (1.52), we find:

$$\dddot{Q} = \begin{bmatrix} 0 & 1 & 0 \\ 0 & 0 & 1 \\ 0 & 0 & 0 \end{bmatrix} \ddot{Q} + \begin{bmatrix} 0 & 0 & 0 \\ 0 & 0 & 0 \\ 0 & 1 & 0 \end{bmatrix} \frac{\partial \Phi}{\partial Q} - \begin{bmatrix} 0 & 0 & 0 \\ 0 & 0 & 0 \\ 0 & 0 & 1 \end{bmatrix} \frac{d}{dt} \frac{\partial \Phi}{\partial Q} + \begin{bmatrix} 0 & 0 & 0 \\ 0 & 0 & 0 \\ 1 & 0 & 0 \end{bmatrix} P. \tag{1.54}$$

From the third row of Eq. (1.54), we can read off the value $p_1 = \dddot{q}_3 - \partial \Phi / \partial x + (d/dt) \partial \Phi / \partial \dot{x}$.

We find that the generalized momentum is:

$$P = \begin{bmatrix} \dddot{q}_3 \\ -\ddot{q}_3 \\ \dot{q}_3 \end{bmatrix} - \begin{bmatrix} 0 & 1 & -d/dt \\ 0 & 0 & 1 \\ 0 & 0 & 0 \end{bmatrix} \frac{\partial \Phi}{\partial Q}. \tag{1.55}$$

Recalling that $q_3 = \ddot{x}$, Eq. (1.55) is equivalent to the form of the generalized momentum in Eq. (1.30).

Finally, Pontryagin's minimum principle (see e.g., [30] Chapter 5) tells us that the Hamiltonian takes a constant value when calculated along an optimal orbit. Comparing Eqs. (1.35) and (1.46), we see that the constant value is the generalized energy $\Psi$, so we find:

$$H = \Psi. \qquad (1.56)$$

This result is, of course, also follows from Eqs. (1.35) and (1.46).

**1.7 Quadratic Φ**

So far, we have only looked at the mechanics of jerk-controlled movements with a Lagrangian containing a general function $\Phi(x,\dot{x},\ddot{x})$. In this section, we look at a class of $\Phi(x,\dot{x},\ddot{x})$ that lead to mathematically straightforward models of movement. Due to their mathematically straightforward form, they are good models to begin with when trying to model a motor behavior. Assuming the jerk-control model is approximately true, for most movements, we expect the model should generate a movement that is a good approximation to the actual movement carried out.

We begin with a general function $\Phi(x,\dot{x},\ddot{x})$. For $\Phi(x,\dot{x},\ddot{x})$ arising in a physical problem, we expect $\Phi(x,\dot{x},\ddot{x})$ is well-behaved, and that we can calculate a Taylor series expansion. We assume that $\Phi(x,\dot{x},\ddot{x})$ can be well-enough approximated by truncating the Taylor series expansion to second-order:

$$\Phi(x,\dot{x},\ddot{x}) \approx \phi_{00} + \begin{bmatrix}\phi_{01}\\ \phi_{02}\\ \phi_{03}\end{bmatrix}^{\mathrm{T}} \begin{bmatrix}x\\ \dot{x}\\ \ddot{x}\end{bmatrix} + \frac{1}{2}\begin{bmatrix}x\\ \dot{x}\\ \ddot{x}\end{bmatrix}^{\mathrm{T}} \begin{bmatrix}\phi_{11} & \phi_{12} & \phi_{13}\\ 0 & \phi_{22} & \phi_{23}\\ 0 & 0 & \phi_{33}\end{bmatrix} \begin{bmatrix}x\\ \dot{x}\\ \ddot{x}\end{bmatrix}. \qquad (1.57)$$

As was the case in Section 1.3 for jerk when we derived the jerk-control cost functional in Eq. (1.6), Eq. (1.57) assumes that the acceleration, velocity, and distance from the origin during the movement remain relatively low. Since $\Phi(x,\dot{x},\ddot{x})$ does not explicitly depend on the time $t$, the elements $\phi_{mn}$ are independent of the time $t$. We can find the optimal orbit models using $\Phi(x,\dot{x},\ddot{x})$ of the form in Eq. (1.57) by solving the Euler-Lagrange; we find:

$$\dddot{x} + \phi_{33}\ddot{x} - (\phi_{22} - \phi_{13})\ddot{x} + \phi_{11}x = -\phi_{01}. \tag{1.58}$$

We observe that a number of the $\phi_{mn}$ that appear in Eq. (1.57) do not appear in Eq. (1.58). This means that the values of these $\phi_{mn}$ do not affect the form of the optimal orbit although they do affect the value of the cost functional $J[\cdot]$. As we do not have a specific physical interpretation of the meaning of the cost functional, and we are only interested in describing the form of the physical movement (i.e. the optimal orbit), we can simplify the model by eliminating a number of the $\phi_{mn}$. We can replace Eq. (1.57) by a more straightforward model for $\Phi(x,\dot{x},\ddot{x})$ that only includes those $\phi_{mn}$ that affect the form of the optimal orbit; we call this the effective model and denote it by $\Phi_{eff}(x,\dot{x},\ddot{x})$. The effective model has the form:

$$\Phi_{eff}(x,\dot{x},\ddot{x}) \approx \phi_{01}x + \frac{1}{2}\begin{bmatrix}x\\ \dot{x}\\ \ddot{x}\end{bmatrix}^T \begin{bmatrix}\phi_{11} & 0 & \phi_{13}\\ 0 & \phi_{22} & 0\\ 0 & 0 & \phi_{33}\end{bmatrix}\begin{bmatrix}x\\ \dot{x}\\ \ddot{x}\end{bmatrix}. \tag{1.59}$$

We note further that $\phi_{13}$ and $\phi_{22}$ do not appear separately in Eq. (1.58), but rather in the difference $\phi_{22} - \phi_{13}$. This means that we are able to arrive at the same optimal orbit with a variety of values of $\phi_{13}$ and $\phi_{22}$; in particular, it means that we can eliminate one of them in Eq. (1.59) if we wish. The natural one to eliminate is $\phi_{13}$ as the meaning of the corresponding term in the Lagrangian is difficult to interpret. Again, both have an effect on the value of the cost functional $J[\cdot]$, and a more physically motivated development of the cost functional would determine which $\phi_{mn}$ must be included in the model.

The generalized energy of the optimal orbit $x(t)$ for quadratic $\Phi(x,\dot{x},\ddot{x})$ is:

$$\Psi = \left(\frac{1}{2}\dddot{x}^2 - \dddot{x}\cdot\ddot{x} + \dddot{x}\cdot\dot{x}\right) - \phi_{33}\cdot\left(\frac{1}{2}\ddot{x}^2 - \dddot{x}\cdot\dot{x}\right)$$
$$-\left(\phi_{22} - \phi_{13}\right)\cdot\left(\frac{1}{2}\dot{x}^2\right) + \left(\phi_{01}x + \frac{1}{2}\phi_{11}x^2\right). \qquad (1.60)$$

We have written the generalized energy as the sum of four terms. We call attention to the pattern in the way the various derivatives of the orbit $x(t)$ appear in each succeeding term of the first three terms on the RHS of Eq. (1.60). Eq. (1.60) is the generalized energy for jerk-control. In the case where the first term on the RHS of Eq. (1.60) is sufficiently small, we can set it to zero in calculating the generalized energy. In that case, the generalized energy takes the form of the generalized energy for acceleration-control. In the case where the first and second terms on the RHS of Eq. (1.60) are sufficiently small, we can set them both to zero, and the generalized energy takes the form of the generalized energy for velocity-control (i.e. the classical energy for classical mechanics).

In practice, this means that if the energy associated with the first term in Eq. (1.60) is sufficiently small, we can treat the problem as approximately an acceleration-control problem. On the other hand, as the pattern will continue for higher orders of control, it is possible that the jerk-control model is just an approximation to a higher-order-control model that better describes motor behavior but for which the generalized energies associated with the (missing) higher order terms in Eq. (1.60) are sufficiently small that we can neglect them.

## 1.8 Remarks on Minimization of Metabolic Work

We have formulated our model of human movement under the assumption that people makes movements that are, in a quantifiable sense, better than the alternatives. We have expressed this mathematically in terms of finding the movement that minimizes a cost functional, but we have left the question of what the cost functional measures open. One reasonable, physical choice for cost functional is the metabolic work. We can think of the metabolic work as the amount of stored fuel (classical energy) that the body must use up to carry out the movement. Under this choice, the person makes the movement that minimizes the metabolic work expended in the course of making the movement. A relationship between the metabolic work of a movement and the integral of the square of the jerk of the movement has been put forward in [15]. We have looked at the application of the optimal jerk-control model developed here where the cost functional is taken to measure the metabolic work in the context of automated exercise coaching in [10].

When the cost functional measures the metabolic work of the movement, we find that the Lagrangian and the Hamiltonian have the units of classical power (i.e. classical energy over time). As a result, the generalized energy $\Psi$ also has units of classical power. In this case, we find that, although the metabolic work is a classical energy, it is not a constant of the motion, but rather accumulates as the person carries out the motion. However, the rate at which the metabolic work accumulates, that is the metabolic power, is related to a constant of the motion given by $\Psi$.

**1.9 Discussion**

We have assumed that a human controls the jerk of a movement and makes the movement in a way that is in some sense better than the available alternatives. From these assumptions, we have constructed a mathematical framework to describe motor behaviors. The framework itself is just a mathematical elaboration of what it means for the movement to better than the alternatives, that is, optimal. What we have presented here is just the mathematical consequences of the physical assumption that the subject carries out the motor behavior by optimal control of the jerk. Confirmation of the empirical accuracy of the mathematical formalism must come through its application to a variety of empirically observed cases of motor behavior.

**1.10 Appendix**

We derive here the existence of the generalized energy $\Psi$ as a constant of the motion for optimal orbits of the optimal jerk-control system using the Lagrangian formalism. Put another way, we derive the law of conservation of generalized energy for such a system.

We begin our derivation of the generalized energy for the jerk-control system by taking the total time derivative of the Lagrangian:

$$\frac{dL}{dt} = \frac{\partial L}{\partial x}\frac{dx}{dt} + \frac{\partial L}{\partial \dot{x}}\frac{d\dot{x}}{dt} + \frac{\partial L}{\partial \ddot{x}}\frac{d\ddot{x}}{dt} + \frac{\partial L}{\partial \dddot{x}}\frac{d\dddot{x}}{dt}. \quad (1.61)$$

The Euler-Lagrange equation (Eq. (1.19)) may be rewritten in the form:

$$\frac{\partial L}{\partial x} = \frac{d}{dt}\frac{\partial L}{\partial \dot{x}} - \frac{d^2}{dt^2}\frac{\partial L}{\partial \ddot{x}} + \frac{d^3}{dt^3}\frac{\partial L}{\partial \dddot{x}}. \quad (1.62)$$

Substituting Eq. (1.62) into Eq. (1.61) gives:

$$\frac{dL}{dt} = \left(\left(\frac{d}{dt}\frac{\partial L}{\partial \dot{x}}\right)\dot{x} + \frac{\partial L}{\partial \dot{x}}\frac{d\dot{x}}{dt}\right) - \left(\left(\frac{d^2}{dt^2}\frac{\partial L}{\partial \ddot{x}}\right)\dot{x} - \frac{\partial L}{\partial \ddot{x}}\frac{d^2\dot{x}}{dt^2}\right) \\ + \left(\frac{d^3}{dt^3}\frac{\partial L}{\partial \dddot{x}}\dot{x} + \frac{\partial L}{\partial \dddot{x}}\frac{d^3\dot{x}}{dt^3}\right). \quad (1.63)$$

We know the forms of several of the derivatives in Eq. (1.63):

$$\frac{d}{dt}\left(\frac{\partial L}{\partial \dot{x}}\dot{x}\right) = \left(\frac{d}{dt}\frac{\partial L}{\partial \dot{x}}\right)\dot{x} + \frac{\partial L}{\partial \dot{x}}\frac{d\dot{x}}{dt},$$

$$\frac{d^2}{dt^2}\left(\frac{\partial L}{\partial \ddot{x}}\dot{x}\right) = \left(\frac{d^2}{dt^2}\frac{\partial L}{\partial \ddot{x}}\right)\dot{x} + 2\left(\frac{d}{dt}\frac{\partial L}{\partial \ddot{x}}\right)\frac{d\dot{x}}{dt} + \frac{\partial L}{\partial \ddot{x}}\frac{d^2\dot{x}}{dt^2}, \quad (1.64)$$

$$\frac{d^3}{dt^3}\left(\frac{\partial L}{\partial \dddot{x}}\dot{x}\right) = \left(\frac{d^3}{dt^3}\frac{\partial L}{\partial \dddot{x}}\right)\dot{x} + 3\left(\frac{d^2}{dt^2}\frac{\partial L}{\partial \dddot{x}}\right)\frac{d\dot{x}}{dt} + 3\left(\frac{d}{dt}\frac{\partial L}{\partial \dddot{x}}\right)\frac{d^2\dot{x}}{dt^2} + \frac{\partial L}{\partial \dddot{x}}\frac{d^3\dot{x}}{dt^3}.$$

Substituting the equations in Eq. (1.64) into Eq. (1.63) gives:

$$\frac{dL}{dt} - \frac{d}{dt}\left(\frac{\partial L}{\partial \dot{x}}\dot{x}\right) + \frac{d^2}{dt^2}\left(\frac{\partial L}{\partial \ddot{x}}\dot{x}\right) - \frac{d^3}{dt^3}\left(\frac{\partial L}{\partial \dddot{x}}\dot{x}\right) =$$
$$2\left(\left(\frac{d}{dt}\frac{\partial L}{\partial \ddot{x}}\right)\frac{d\dot{x}}{dt} + \frac{\partial L}{\partial \ddot{x}}\frac{d^2\dot{x}}{dt^2}\right) - 3\left(\left(\frac{d^2}{dt^2}\frac{\partial L}{\partial \dddot{x}}\right)\frac{d\dot{x}}{dt} + \left(\frac{d}{dt}\frac{\partial L}{\partial \dddot{x}}\right)\frac{d^2\dot{x}}{dt^2}\right).$$
(1.65)

We may rewrite Eq. (1.65) in the form:

$$\frac{dL}{dt} - \frac{d}{dt}\left(\frac{\partial L}{\partial \dot{x}}\dot{x}\right) + \frac{d^2}{dt^2}\left(\frac{\partial L}{\partial \ddot{x}}\dot{x}\right) - \frac{d^3}{dt^3}\left(\frac{\partial L}{\partial \dddot{x}}\dot{x}\right) =$$
$$2\left(\left(\frac{d}{dt}\frac{\partial L}{\partial \ddot{x}}\right)\ddot{x} + \frac{\partial L}{\partial \ddot{x}}\frac{d\ddot{x}}{dt}\right) - 3\left(\left(\frac{d^2}{dt^2}\frac{\partial L}{\partial \dddot{x}}\right)\ddot{x} + \left(\frac{d}{dt}\frac{\partial L}{\partial \dddot{x}}\right)\frac{d\ddot{x}}{dt}\right).$$
(1.66)

Again, we know the forms of several of the derivatives in Eq. (1.66):

$$\frac{d}{dt}\left(\frac{\partial L}{\partial \ddot{x}}\ddot{x}\right) = \left(\frac{d}{dt}\frac{\partial L}{\partial \ddot{x}}\right)\ddot{x} + \frac{\partial L}{\partial \ddot{x}}\frac{d\ddot{x}}{dt},$$

$$\frac{d^2}{dt^2}\left(\frac{\partial L}{\partial \dddot{x}}\ddot{x}\right) = \left(\frac{d^2}{dt^2}\frac{\partial L}{\partial \dddot{x}}\right)\ddot{x} + 2\left(\frac{d}{dt}\frac{\partial L}{\partial \dddot{x}}\right)\frac{d\ddot{x}}{dt} + \frac{\partial L}{\partial \dddot{x}}\frac{d^2\ddot{x}}{dt^2}.$$
(1.67)

Again, substituting the equations in Eq. (1.67) into Eq. (1.67) gives:

$$\frac{dL}{dt} - \frac{d}{dt}\left(\frac{\partial L}{\partial \dot{x}}\dot{x}\right) + \frac{d^2}{dt^2}\left(\frac{\partial L}{\partial \ddot{x}}\dot{x}\right) - \frac{d^3}{dt^3}\left(\frac{\partial L}{\partial \dddot{x}}\dot{x}\right)$$
$$-2\frac{d}{dt}\left(\frac{\partial L}{\partial \ddot{x}}\ddot{x}\right) + 3\frac{d^2}{dt^2}\left(\frac{\partial L}{\partial \dddot{x}}\ddot{x}\right) = 3\left(\left(\frac{d}{dt}\frac{\partial L}{\partial \dddot{x}}\right)\frac{d\ddot{x}}{dt} + \frac{\partial L}{\partial \dddot{x}}\frac{d^2\ddot{x}}{dt^2}\right).$$
(1.68)

We may rewrite Eq. (1.68) as:

$$\frac{d}{dt}\left(\frac{d^2}{dt^2}\left(\frac{\partial L}{\partial \dddot{x}}\dot{x}\right) - \frac{d}{dt}\left(\frac{\partial L}{\partial \ddot{x}}\dot{x}\right) - 3\frac{d}{dt}\left(\frac{\partial L}{\partial \dddot{x}}\ddot{x}\right)\right.$$
$$\left. + \frac{\partial L}{\partial \dot{x}}\dot{x} + 2\frac{\partial L}{\partial \ddot{x}}\ddot{x} + 3\frac{\partial L}{\partial \dddot{x}}\dddot{x} - L\right) = 0.$$
(1.69)

Eq. (1.69) indicates that the time-derivative of a quantity is equal to zero, so that quantity is a constant of the motion; we write this as:

$$\frac{d^2}{dt^2}\left(\frac{\partial L}{\partial \dddot{x}}\dot{x}\right) - \frac{d}{dt}\left(\frac{\partial L}{\partial \ddot{x}}\dot{x}\right) - 3\frac{d}{dt}\left(\frac{\partial L}{\partial \dddot{x}}\ddot{x}\right) + \frac{\partial L}{\partial \dot{x}}\dot{x} + 2\frac{\partial L}{\partial \ddot{x}}\ddot{x} + 3\frac{\partial L}{\partial \dddot{x}}\dddot{x} - L = \Psi. \quad (1.70)$$

We can continue and rewrite the LHS of Eq. (1.70) in a more compact form. As before, we know the forms of several of the derivatives in Eq. (1.70):

$$\frac{d^2}{dt^2}\left(\frac{\partial L}{\partial \ddot{x}}\dot{x}\right) = \left(\frac{d^2}{dt^2}\frac{\partial L}{\partial \ddot{x}}\right)\dot{x} + 2\left(\frac{d}{dt}\frac{\partial L}{\partial \ddot{x}}\right)\ddot{x} + \frac{\partial L}{\partial \ddot{x}}\dddot{x},$$

$$\frac{d}{dt}\left(\frac{\partial L}{\partial \dot{x}}\dot{x}\right) = \left(\frac{d}{dt}\frac{\partial L}{\partial \dot{x}}\right)\dot{x} + \frac{\partial L}{\partial \dot{x}}\ddot{x}, \quad (1.71)$$

$$\frac{d}{dt}\left(\frac{\partial L}{\partial \ddot{x}}\ddot{x}\right) = \left(\frac{d}{dt}\frac{\partial L}{\partial \ddot{x}}\right)\ddot{x} + \frac{\partial L}{\partial \ddot{x}}\dddot{x}.$$

As before, substituting the equations in Eq. (1.71) into Eq. (1.70) gives:

$$\left(\frac{d^2}{dt^2}\frac{\partial L}{\partial \ddot{x}} - \frac{d}{dt}\frac{\partial L}{\partial \dot{x}} + \frac{\partial L}{\partial x}\right)\dot{x} + \left(-\frac{d}{dt}\frac{\partial L}{\partial \ddot{x}} + \frac{\partial L}{\partial \dot{x}}\right)\ddot{x} + \frac{\partial L}{\partial \ddot{x}}\dddot{x} - L = \Psi. \quad (1.72)$$

Using the generalized momentum given in Eq. (1.30), we may rewrite Eq. (1.72) in the compact form:

$$P^{\mathrm{T}}\dot{X} - L = \Psi. \quad (1.73)$$

The generalized energy is the constant of the motion $\Psi$. We observe that the quantity $P^{\mathrm{T}}\dot{X} - L$ is conserved through an orbit that optimizes the cost functional.

# Chapter 2 – On the Description of Rapid, Targeted Movements of a Computer Mouse

**Abstract.** We use the optimal jerk-control model of human movements developed in Chapter 1 to formulate a mathematical model describing how a subject makes rapid, targeted movements of a computer mouse some into a defined target region. These movements are known to be described mathematically by Fitts' law. In the model we developed in Chapter 1, the subject selects total movement time prior to beginning the movement, a model that seems to run counter to intuitive experience. We show that we can instead formulate the model so that the subject selects the generalized energy prior to beginning the movement and that the total movement time follows from that selection. We use this approach to formulating the model to show that, using a particular choice of the generalized energy, we can make the mouse movements described by the model satisfy Fitts' law.

## 2.1 Introduction

When using a computer mouse to control a computer, a subject repeats many times the straightforward task of making a rapid movement that must end in a designated target region. The task of moving the mouse is an example of a speed/accuracy trade off in which the subject must weigh the goal of completing the task as quickly as possible against the likelihood of missing the target (i.e. ending the movement outside of the target region). Thus, in each movement of the mouse, the subject exhibits their cognitive and motor proficiency at controlling the mouse using the hand as well as their willingness to make mistakes when using the mouse (i.e. miss the target region). As a result, mouse movements provide a potentially rich source of information about a subject that can be measured regularly and frequently in the home. In this vein, we have proposed the use of measurements of computer mouse usage for early detection of decline in cognitive performance in older adults to help provide appropriate medical care. Two methods that have been proposed are: (1) measuring computer mouse movements made in the course of playing a specifically designed computer game, [10] and (2) making measurements from everyday computer mouse usage. [11] In this Chapter, we use the optimal jerk-control model of basic human movements to develop a model to describe how a subject moves a mouse. The aim is to develop a more detailed framework for understanding mouse movements than those used in the two methods and use it to understand observed mouse usage better.

To develop the desired framework for understanding mouse movements using the optimal jerk-control model, we have to look at how a subject makes mouse movements and how to make the model best describe that process. As we have said, a computer mouse movement is a rapid movement into a target region. This means that, once the subject has decided on the target (i.e. the widget that the subject wishes to select) the position of the target and the size of the target region are given. The starting point of the mouse is outside the subject's control since the subject starts moving the mouse from wherever it happens to be when the subject decides to move the mouse to the target. We measure the time it takes the subject to move to the target and whether the subject hits or misses the target (i.e. whether the movement ends in the target region). This suggests that the correct way of understanding mouse movements is to see the total movement time as arising from the constraints on the specific movement that the subject is carrying out. However, in the formulation of the optimal jerk-control model in Chapter 1, the subject chooses movement time prior to beginning the movement and then just carries out the movement that takes the chosen amount of time. In this Chapter, we show that we can reformulate the optimal jerk-control model so that the subject chooses the generalized energy of the movement prior to beginning the movement and that the total movement time arises from the constraints in the movement problem and the choice of the generalized energy. We argue that the value of the generalized energy corresponds to a subject's subjective impression of how "hard," or "strong," or "fast" a mouse movement is.

Having shown that we can reformulate the optimal jerk-control movement problem as a problem in which the subject chooses the generalized energy rather than the total movement time prior to making the making the movement, we show that the resulting model can be made to approximate the standard model for movements with a speed/accuracy trade off – Fitts' law. We do this by choosing a generalized energy that is proportional to the square of the width of the target region along the line of the mouse movement. We conclude by examining alternative optimal control models and seeing if the alternative controls yield better approximations to Fitts' law than does the choice of jerk-control.

We note that Fitts' law provides the summary measure of mouse usage used in the method in [10]. Thus, in providing an approximation to Fitts' law, the optimal jerk-control model developed here informs what those methods of in-home monitoring using observed movements of the mouse are measuring.

**2.2 Fitts' Law**

Movements of a mouse in the course of using a computer are an example of rapid, targeted movements [31, 32] which are described by Fitts' law. [33-35] As Fitts' law is an empirically established measure, any physically correct model describing movements of the mouse should approximately satisfy Fitts' law in the situations in which Fitts' law has been observed to hold. Fitts' law is an empirical formula that fits observed rapid, target movements well. The form of Fitts' law that we use is taken from [35]; though alternative, slightly different, formulas also have been proposed (see e.g., [33, 36]). A comparison of the performance of the formulations of Fitts' law given in [33, 35, 36] using empirical data is provided in [35]. Applications of Fitts' law have been made to the design of graphical user interfaces [31] and to capturing the contribution of physical movement time in the play of a computer game. [10]

A subject navigates a computer using a mouse by physically moving the mouse. One way a subject uses the mouse is by moving the mouse causing the pointer on the computer screen to move some distance from a starting position into a target region or *icon* on the screen that corresponds to a control widget (e.g., a button, a scrollbar, etc.). The subject then presses a button on the mouse to select the control widget. Another way a subject uses the mouse is by selecting an icon as before and then, holding the mouse button down, the subject *drags* the icon to a new location on the computer screen.

When the subject moves the mouse to take the pointer into a target region, the subject is making a rapid, targeted movement. In a rapid, targeted movement, the subject must make a movement into the target region as quickly as possible while being sure that the movement ends in the target region with sufficient reliability. Thus, a subject can make quicker movements that miss the target more often, or slower movements that miss the target less often. Fitts' law provides a mathematical relationship between the distance $D$ moved from the initial position to the center of the target region, the width $W$ of the target region, and the total movement time $T$ required to complete the movement. In Fitts' law, these three quantities are related according to the formula:

$$T = a + b \log_2(D/W + 1). \tag{2.1}$$

We can think of the physical movement (the hand on the mouse in the case of mouse movements) as happening in conjunction with cognitive process that looks at how the movement that adjusts the motion of the hand as the movement proceeds. The *index of difficulty* $\log_2(D/W+1)$ provides a measurement of the amount of information that must be processed in the parallel cognitive process; as the amount of information to be processed increases, the movement slows (i.e. the movement time increases) to accommodate the extra time needed to process the additional information; it is measured in bits. The constant $b$ gives the time the cognitive process takes to process each bit of information. As Fitts' law describes a movement that starts outside the target region, the lower bound on the distance moved is $D_{LB} = W/2$. We can use this to derive a lower bound for the total movement time:

$$T_{LB} = a + b\log_2(3/2). \tag{2.2}$$

Empirical measurements have been made of typical values of the parameters $a$ and $b$ in Eq. (2.1) for three methods of deleting an icon on an Apple Macintosh computer. [32] The first method was *point-select*; subjects carried out point-select deletion by first selecting the icon by clicking on it and then deleting the icon by selecting the trashcan by clicking on it. The second method was *drag-select*; subjects carried out drag-select deletion by clicking on the icon, but instead of the releasing the button, holding the button down and dragging the icon to the trashcan. The third method was *stroke-through*; subjects carried out stroke-through deletion by bringing the pointer to one side of the icon, hold the button down and stroke through the icon, and finally release the mouse button on the other side of the icon. In the case of point-select movements, Fitts' law was found to take the form:

$$T = [230\,\text{ms}] + [166\,\text{ms/bit}]\log_2(D/W+1). \tag{2.3}$$

In the case of drag-select and stroke-through movements, Fitts' law was found to take the form:

$$T = [135\,\text{ms}] + [249\,\text{ms/bit}]\log_2(D/W + 1). \tag{2.4}$$

Eqs. (2.3) and (2.4) actually yield similar values for the lower bound on the total movement time with the lower bound for point-select movements being $T_{LB}$ = 330 ms and the lower bound for drag-select and stroke-through movements being $T_{LB}$ = 280 ms.

The different values in Eqs. (2.3) and (2.4) come from whether subjects hold down the mouse button during the movement. If we were to approximate the two equations so that they had the same lower bound on the movement time, then the mathematical difference between the two equations would come down to the amount of time required to process each bit of information expressed in $b$. It appears that, as the movement distance increases, the amount of time the subject requires for processing the information increases faster for movements in which the subject holds mouse button down than in movements in which the subject does not.

## 2.3 Optimal Jerk-Control of Mouse Movements

In Chapter 1, we developed the formalism for modeling motor behaviors using optimal control of the jerk. We now construct, using this formalism, a model for describing the rapid, targeted movements of a mouse.

In the case of the mouse, it is important to distinguish two spaces in which the movement takes place: (1) mouse space, and (2) pointer space. Mouse space is the physical space of the table on which the hand moves the mouse, while pointer space is the virtual space of the computer screen in which the pointer moves. Movement of the pointer in pointer space is related to movement of the mouse in mouse space by a (generally non-linear) transfer function (see e.g. the transfer function for the Windows XP operating system as described in [37]). To keep the present analysis of mouse movements straightforward, we will assume that, to good approximation, we can restrict our analysis to looking only at the mouse movements as they occur in pointer space.

Thus, in this analysis, we look at how the pointer moves on the computer screen rather than how the hand moves the mouse on the table.

We assume that the target region is centered at the origin and has width $W$, that the movement begins at a position $-D$ and ends at the center of the target region, and that the movement takes some total time $T$. Although the formalism we employ here requires that we assume knowledge of the movement time $T$ prior to beginning the movement, we show in Section 2.4, we can replace this assumption with an assumption about the value of another variable, the generalized energy $\Psi$ and leave the actual movement time unknown at the beginning of the movement.

We treat a mouse movement as a basic movement (i.e. a simple movement that begins and ends at rest) whose orbit $x(t)$ is in some sense better than any alternative orbit satisfying the same conditions; that is, the movement is optimal in the sense of minimizing some measure of cost that we can calculate for all possible movements. Mathematically, we formulate the problem of finding this optimal orbit as that of finding the movement that minimizes a jerk-control cost functional $J[\cdot]$ subject to the stated initial and final conditions; that is:

$$J[x] = \int_0^T L(x, \dot{x}, \ddot{x}, \dddot{x}) dt,$$
$$x(0) = -D, \quad x(T) = 0,$$
$$\dot{x}(0) = 0, \quad \dot{x}(T) = 0, \qquad (2.5)$$
$$\ddot{x}(0) = 0, \quad \ddot{x}(T) = 0.$$

We have implicitly assumed that the Lagrangian $L(x, \dot{x}, \ddot{x}, \dddot{x})$ in Eq. (2.5) has no explicit time-dependence, so that the rate of change in the cost functional at an instant $t$ depends only on the instantaneous state of the orbit $x(t)$. We have also assumed that Lagrangian in Eq. (2.5) is independent of the total movement time $T$, that is, that the instantaneous mechanics of the movement does not depend on the amount of time the movement will ultimately take. Assuming (as in Chapter 1) that there is no interaction between the cost associated with the jerk (the control) and the position, velocity, or acceleration, then (following Chapter 1), we can write the Lagrangian for Eq. (2.5) in the form:

$$L = \frac{1}{2}\dddot{x}^2 - \Phi(x, \dot{x}, \ddot{x}). \tag{2.6}$$

Our aim is to use a jerk-control model that we can express in the form of Eq. (2.5) to describe basic movements of a mouse. Although we do not know the physical form of the Lagrangian in Eq. (2.5), we assume, following Chapter 1, that we can approximate $\Phi(x, \dot{x}, \ddot{x})$ using a second-order Taylor series expansion. As we are only interested in finding the orbit $x(t)$ that minimizes the cost functional, we, again following Chapter 1, further simplify by using the effective model $\Phi_{eff}(x, \dot{x}, \ddot{x})$ that only contains those parts of $\Phi(x, \dot{x}, \ddot{x})$ that affect the form of the optimal orbit. We replace Eq. (2.6) with an effective Lagrangian of the form:

$$L_{eff} = \frac{1}{2}\dddot{x}^2 - \phi_{01}x - \frac{1}{2}\begin{bmatrix} x \\ \dot{x} \\ \ddot{x} \end{bmatrix}^T \begin{bmatrix} \phi_{11} & 0 & \phi_{13} \\ 0 & \phi_{22} & 0 \\ 0 & 0 & \phi_{33} \end{bmatrix} \begin{bmatrix} x \\ \dot{x} \\ \ddot{x} \end{bmatrix}. \tag{2.7}$$

Again, the elements $\phi_{mn}$ are independent of both the time $t$ and the total movement time $T$.

The portion of the effective Lagrangian (Eq. (2.7)) that depends only on the instantaneous position in space during the movement is:

$$L'_{eff}(x) = -\phi_{01}x - \frac{1}{2}\phi_{11}x^2. \tag{2.8}$$

This portion of the effective Lagrangian describes the cost associated with occupying various positions in space. The goal of the mouse movement is to end the movement within the target region. As ending the movement at any point within the target region meets this goal, we would like to have the cost associated with the target region be approximately constant, we do this by requiring:

$$\frac{dL'_{eff}}{dx}(0) = 0. \tag{2.9}$$

This requires we take $\phi_{01} = 0$; the effective Lagrangian becomes:

$$L_{eff} = \frac{1}{2}\ddot{x}^2 - \frac{1}{2}\begin{bmatrix} x \\ \dot{x} \\ \ddot{x} \end{bmatrix}^T \begin{bmatrix} \phi_{11} & 0 & \phi_{13} \\ 0 & \phi_{22} & 0 \\ 0 & 0 & \phi_{33} \end{bmatrix} \begin{bmatrix} x \\ \dot{x} \\ \ddot{x} \end{bmatrix}. \tag{2.10}$$

We note that the Lagrangian in Eq. (2.10) looks much like a generalization of the Lagrangian for the simple harmonic oscillator in classical mechanics.

We calculate the optimal orbit $x(t)$ by finding the orbit that minimizes the cost functional $J[\cdot]$ in Eq. (2.5); we do this by setting the first variation of $J[\cdot]$ to zero. Thus, following Chapter 1, the optimal orbit satisfies:

$$\frac{d^3}{dt^3}\frac{\partial L}{\partial \ddot{x}} - \frac{d^2}{dt^2}\frac{\partial L}{\partial \ddot{x}} + \frac{d}{dt}\frac{\partial L}{\partial \dot{x}} - \frac{\partial L}{\partial x} = 0. \tag{2.11}$$

Applying Eq. (2.11) to Eq. (2.10) gives the Euler-Lagrange equation for the optimal orbit $x(t)$:

$$\ddddot{x} + \phi_{33}\dddot{x} - (\phi_{22} - \phi_{13})\ddot{x} + \phi_{11}x = 0. \tag{2.12}$$

The subject makes the movement in such a way as to satisfy Eq. (2.12) as well as the initial and final conditions given in Eq. (2.5).

From the effective Lagrangian in Eq. (2.10), we can calculate a dual quantity called the *Hamiltonian* by applying the Legendre transform [29] to the effective Lagrangian (see Chapter 1). The Hamiltonian takes the form:

$$H = \frac{1}{2}P^{\mathrm{T}}\begin{bmatrix} 0 & 0 & 0 \\ 0 & 0 & 0 \\ 0 & 0 & 1 \end{bmatrix}P + P^{\mathrm{T}}\begin{bmatrix} 0 & 1 & 0 \\ 0 & 0 & 1 \\ 0 & 0 & 0 \end{bmatrix}Q + \frac{1}{2}Q^{\mathrm{T}}\begin{bmatrix} \phi_{11} & 0 & \phi_{13} \\ 0 & \phi_{22} & 0 \\ 0 & 0 & \phi_{33} \end{bmatrix}Q. \qquad (2.13)$$

In this case, we have vectors $Q^{\mathrm{T}} = \begin{bmatrix} q_1 & q_2 & q_3 \end{bmatrix}$ and $P^{\mathrm{T}} = \begin{bmatrix} p_1 & p_2 & p_3 \end{bmatrix}$ that satisfy the equation (see Chapter 1):

$$\dot{Q} = \begin{bmatrix} 0 & 1 & 0 \\ 0 & 0 & 1 \\ 0 & 0 & 0 \end{bmatrix}Q + \begin{bmatrix} 0 & 0 & 0 \\ 0 & 0 & 0 \\ 0 & 0 & 1 \end{bmatrix}P. \qquad (2.14)$$

According to Pontryagin's minimum principle (see e.g., [30] Chapter 5), the Hamiltonian takes a constant value when calculated along an optimal orbit that minimizes the cost functional calculated from the associated Lagrangian. This constant is the generalized energy $\Psi$; so we find that, along an optimal orbit (i.e. an orbit that satisfies Eq. (2.12) where $q_1(t) = x(t)$), we have:

$$H = \Psi. \qquad (2.15)$$

As the Hamiltonian equals the generalized energy $\Psi$ over the entire optimal orbit, it must equal it at time $t = 0$ at which time the generalized coordinates and momentum satisfy the initial condition in Eq. (2.5); that is:

$$Q(0) = \begin{bmatrix} -D \\ 0 \\ 0 \end{bmatrix}, \quad P(0) = \begin{bmatrix} p_1 \\ p_2 \\ \ddot{x}(0) \end{bmatrix}. \qquad (2.16)$$

Evaluating Eqs. (2.13) and (2.15) at time $t = 0$, we find:

$$\frac{1}{2}\ddot{x}(0)^2 + \frac{1}{2}\phi_{11}D^2 = \Psi. \qquad (2.17)$$

If we assume that the values $\phi_{11}T^6$ and $\phi_{33}T^2$ are small (assumptions that appear to be supported by the results given in Section 2.4), then initial value of the jerk is approximately (see Section 2.7.1):

$$\dddot{x}(0) \approx \left(60 - 24\phi_{33}T^2 - \frac{\phi_{11}T^6}{120} + \frac{54\phi_{33}^2 T^4}{5} + \frac{9\phi_{11}\phi_{33}T^8}{20}\right)\frac{D}{T^3}. \qquad (2.18)$$

Plugging Eq. (2.18) into Eq. (2.17) gives:

$$\frac{1}{2}\left(60 - 24\phi_{33}T^2 - \frac{\phi_{11}T^6}{120} + \frac{54\phi_{33}^2 T^4}{5} + \frac{9\phi_{11}\phi_{33}T^8}{20}\right)^2 \frac{D^2}{T^6} + \frac{1}{2}\phi_{11}D^2 \approx \Psi. \qquad (2.19)$$

Evaluating the leading term on the LHS of Eq. (2.19) and keeping to at most second-order in $\phi_{11}T^6$ and $\phi_{33}T^2$, we find:

$$1800 - 1440\phi_{33}T^2 + 816\phi_{33}^2 T^4 - \frac{\Psi}{D^2}T^6 + 27\phi_{11}\phi_{33}T^8 + \frac{\phi_{11}^2}{28800}T^{12} \approx 0. \qquad (2.20)$$

Eq. (2.20) is a sixth-order polynomial in $T^2$ expressing a relationship between the total distance moved $D$, total movement time $T$, the unknown parameters $\phi_{11}$ and $\phi_{33}$, and the generalized energy $\Psi$. As it relates the distance moved to the movement time, we expect Eq. (2.20) to agree with Fitts' law (Eq. (2.1)) when applied to rapid, targeted mouse movements.

## 2.4 Optimally Jerk-Controlled Mouse Movements & Fitts' Law

In Section 2.3, we formulated the mouse movement problem as one of finding the orbit $x(t)$ that minimizes the cost functional $J[\cdot]$ while a total distance $D$ into a specified target region of width $W$ in a specified total movement time $T$, while beginning and ending the movement at rest. We used the resulting optimal orbit $x(t)$ to calculate the generalized energy $\Psi$ associated with the movement. However, this formulation of the problem does not seem to correspond to intuitive experience. We do not seem in most cases to make movements knowing the total movement time in advance. Instead, we seem to have a sense of one movement being "harder," "stronger," or "faster" than another. We would like to reformulate the model developed in Section 2.3 so that we specify the "hardness" rather than the total movement time $T$.

The only free parameters available in the model for representing "hardness" are the parameters $\phi_{mn}$ and the generalized energy $\Psi$. The intent of the parameters $\phi_{mn}$ is to provide a description of the cost functional $J[\cdot]$ that the movement is minimizing. We expect to be able to find movements of different "hardness" to satisfy the same cost functional, but by doing so using different movement times. The generalized energy, on the other hand, describes the entire movement and changes value when the total movement time $T$ changes. Thus, there is a kind of duality between the generalized energy and the total movement time in the model. As a result, we argue that the natural parameter to use for expressing the "hardness" of a movement in the formalism that we have developed is the generalized energy $\Psi$.

*2.4.1 Analogy with Classical Mechanics*

We motivate the reformulation of the movement into one in which the generalized energy $\Psi$ is specified at the beginning of the movement rather than the total movement time $T$ by looking at the analogous problem in classical mechanics. In doing this, we aim to show how the choice of the (conserved) classical energy $E$ determines the movement time $T$.

A classical particle of mass $m$ travels a distance $D$ in time $T$ in free space with no external interactions moves in an orbit $x(t)$ that optimizes the action, that is, cost functional given by:

$$S[x] = \int_0^T L(x,\dot{x})dt = \int_0^T \frac{1}{2}m\dot{x}^2 dt. \qquad (2.21)$$

The Hamiltonian for the system described by Eq. (2.21) is:

$$H = \frac{1}{2}m\dot{x}^2. \tag{2.22}$$

When evaluated on an orbit $x(t)$ that optimizes the cost functional in Eq. (2.21), the Hamiltonian has a constant value; in this case, that value is the classical energy $E$, so:

$$\frac{1}{2}m\dot{x}^2 = E. \tag{2.23}$$

As there is no force acting on the particle, the velocity takes on the constant value $\dot{x} = D/T$, so we arrive at the expression:

$$\frac{D^2}{T^2} = \frac{2E}{m}. \tag{2.24}$$

By solving Eq. (2.24) for the time $T$, we find the time needed for the particle to move the given distance given that it begins with a classical energy $E$:

$$T = \frac{1}{\sqrt{2}} \frac{D}{\sqrt{E/m}}. \tag{2.25}$$

We see, that for higher classical energies $E$, the particle traverses the required distance $D$ in shorter times $T$. We also note that the value of the classical energy uniquely determines the time that the movement takes. Indeed, in classical mechanics, it is perfectly sensible to pose the problem of finding the time needed to complete the movement given that the particle begins with a specified energy (and thus a specified velocity).

*2.4.2 Case 1: $\phi_{11} = \phi_{33} = 0$*

We now look at Eq. (2.20) in the most straightforward case, that is, the case where $\phi_{11} = \phi_{33} = 0$. In this case, we can rewrite Eq. (2.20) in the form:

$$T^6 \approx 1800 \frac{D^2}{\Psi}. \tag{2.26}$$

We can see that, as is the case in Eq. (2.25), in Eq. (2.26) the value of the generalized energy $\Psi$ determines to total movement time $T$.

We note that, for reasonable values of $\xi$ and appropriate values of $\tau$, $a$, and $b$, we can make the following approximation:

$$\tau \xi^{1/3} \approx a + b \log_2(\xi + 1). \qquad (2.27)$$

This means that Eq. (2.26) approximates Fitts' law when we choose the generalized energy $\Psi$ proportional to the square of width of the target region $W$, that is:

$$\Psi = \frac{1800}{\tau^6} W^2. \qquad (2.28)$$

Plugging Eq. (2.28) into Eq. (2.26), we find that we arrive at Fitts' law:

$$T \approx \tau \cdot \left(\frac{D}{W}\right)^{1/3}. \qquad (2.29)$$

In Eq. (2.27), the LHS has one free parameter while the RHS has two, so it is possible that for empirical measurements of Fitts' law, we might not be able to approximate the observed form of Fitts' law in Eq. (2.1) using the model in Eq. (2.29). We now look at how well the model for Fitts' law in Eq. (2.29) describes observed mouse movements by compare the formulations for Fitts' law in Eq. (2.29) to that in Eq. (2.1) using the measured values for $a$ and $b$ in [32]. We evaluate the two formulations on the range $1 \leq D/W \leq 64$ corresponding to the range on which mouse movements were measured in [32]. We proceed by finding the parameter value $\tau$ that provides the best approximation to the formulation in $a$ and $b$ by fitting Eq. (2.29) to Eq. (2.1) using ordinary least squares (OLS) regression on the indicated range using a evenly spaced sample of values on $D/W$.

We begin by looking at the point-select movements described by Eq. (2.3), that is, movements where the subject *does not* hold down the mouse button during the movement. In this case, the problem is to find the value of $\tau$ that provides the best approximation to Eq. (2.3); that is:

$$\tau \cdot \left(\frac{D}{W}\right)^{1/3} \approx [230\,\text{ms}] + [166\,\text{ms/bit}]\log_2\left(\frac{D}{W}+1\right). \tag{2.30}$$

Performing the OLS regression over the range $1 \leq D/W \leq 64$, we find:

$$\tau = 330\,\text{ms}. \tag{2.31}$$

The corresponding lower bound on the total movement time is $T_{LB} = 260$ ms. We provide a comparison of the functional form of the model in Eq. (2.29) to Eq. (2.3) for this case in Figure 2.1.

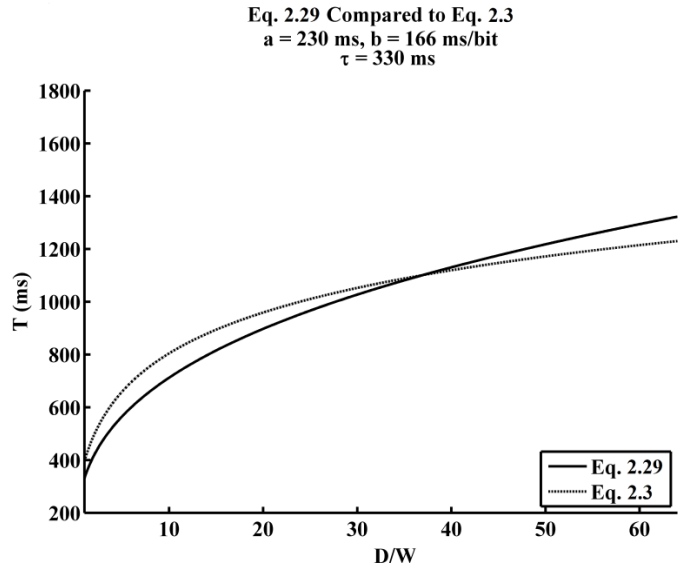

**Figure 2.1.** Eq. (2.29) compared to Eq. (2.3). The model in Eq. (2.29) is compared to the empirical formulation of Fitts' law in Eq. (2.3) for the case of $a = 230$ ms and $b = 166$ ms/bit over the range $1 \leq D/W \leq 64$.

Moving on to drag-select and stroke-through movements, where the subject *does* hold down the mouse button during the movement. In this case, the problem is to find the value of $\tau$ that provides the best approximation to Eq. (2.4); that is:

$$\tau \cdot \left(\frac{D}{W}\right)^{1/3} \approx [135\,\text{ms}] + [249\,\text{ms/bit}]\log_2\left(\frac{D}{W}+1\right). \tag{2.32}$$

Performing the OLS regression over the range $1 \leq D/W \leq 64$, we find:

$$\tau = 430\,\text{ms}. \tag{2.33}$$

The corresponding lower bound on the total movement time is $T_{LB} = 340$ ms. We provide a comparison of the functional form of the model in Eq. (2.29) to Eq. (2.4) for this case in Figure 2.2.

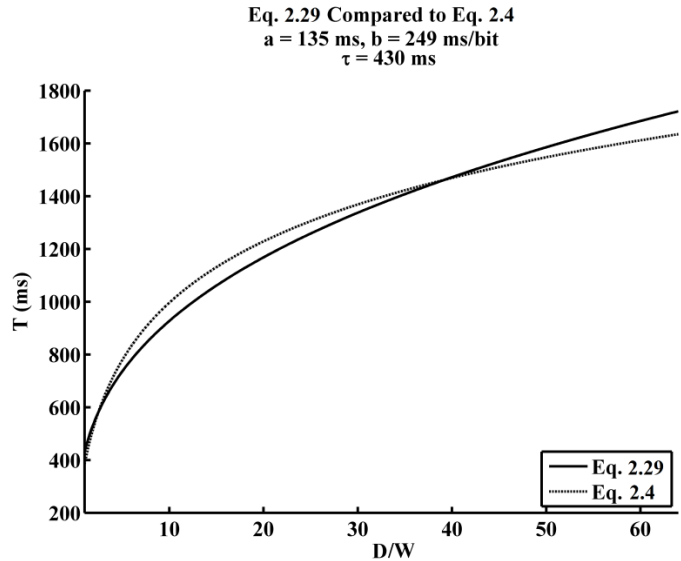

**Figure 2.2.** Eq. (2.29) compared to Eq. (2.4). The model in Eq. (2.29) is compared to the empirical formulation of Fitts' law in Eq. (2.4) for the case of $a = 135$ ms and $b = 249$ ms/bit over the range $1 \leq D/W \leq 64$.

### 2.4.3 Case 2: $\phi_{33} = 0$

We next look at the somewhat more complicated example in which we set $\phi_{33} = 0$ in Eq. (2.20). In constructing the model in Section 2.3, we have required that $\phi_{11}$ be a constant that is independent of the movement time $T$. Having shown that the generalized energy $\Psi$ determines the movement time, we extend that requirement to having $\phi_{11}$ be independent of $\Psi$ as well.

In the case of $\phi_{33} = 0$, we can rewrite Eq. (2.20) in the form:

$$\frac{\phi_{11}^2}{4 \cdot 60^4} T^{12} - \tau^{-6} \cdot \left(\frac{W}{D}\right)^2 T^6 + 1 \approx 0. \tag{2.34}$$

We note that Eq. (2.34) is quadratic in $T^6$. An approximate solution to Eq. (2.34) that is suitable for estimating the value of $\phi_{11}$ given values of $D/W$ and $T$ is (see Section 2.7.2):

$$T \approx \tau \cdot \left(\frac{D}{W}\right)^{1/3} \left(1 - \frac{(\phi_{11}\tau^6)^2}{24 \cdot 60^4} \left(\frac{D}{W}\right)^4\right). \tag{2.35}$$

We can look at Eq. (2.35) as being the sum of Eq. (2.29) and a corrective term of second-order in $\phi_{11}$.

In the case of point-select movements, where the subject *does not* hold down the mouse during the movement, using the Fitts' law from Eq. (2.3), we find:

$$\tau \cdot \left(\frac{D}{W}\right)^{1/3} - \frac{\tau \cdot (\phi_{11}\tau^6)^2}{24 \cdot 60^4} \cdot \left(\frac{D}{W}\right)^{13/3} \approx [230\,\text{ms}] + [166\,\text{ms/bit}] \log_2\left(\frac{D}{W} + 1\right). \tag{2.36}$$

Performing the OLS regression over the range $1 \leq D/W \leq 64$, we find:

$$\begin{aligned} \tau &\approx 350\,\text{ms}, \\ |\phi_{11}|\tau^6 &\approx 1.7. \end{aligned} \tag{2.37}$$

The corresponding lower bound on the total movement time is $T_{LB}$ = 280 ms. We provide a comparison of the functional form of the model in Eq. (2.35) to Eq. (2.3) for this case in Figure 2.3. We note that the approximations used to derive Eq. (2.35) break down for $D/W$ greater than about 50.

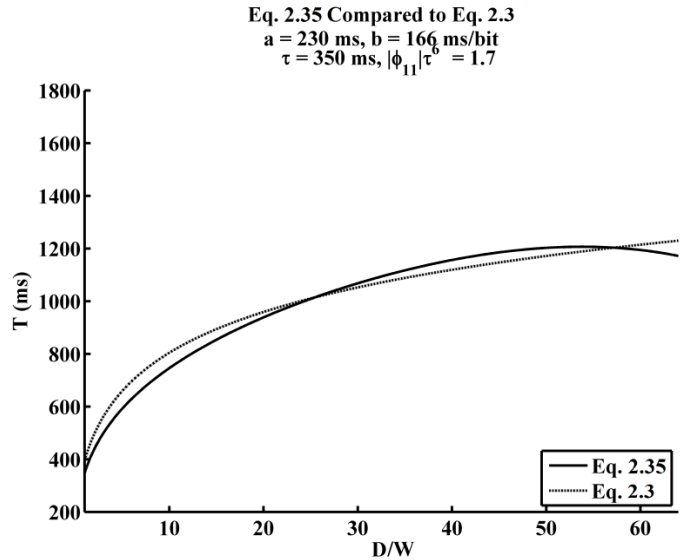

**Figure 2.3.** Eq. (2.35) compared to Eq. (2.3). The model in Eq. (2.35) is compared to the empirical formulation of Fitts' law in Eq. (2.3) for the case of $a$ = 230 ms and $b$ = 166 ms/bit over the range $1 \leq D/W \leq 64$.

In the case of drag-select and stroke-through movements, where the subject *does* hold down the mouse button during the movement, using the Fitts' law from Eq. (2.4), we find:

$$\tau \cdot \left(\frac{D}{W}\right)^{1/3} - \frac{\tau \cdot (\phi_{11}\tau^6)^2}{24 \cdot 60^4} \cdot \left(\frac{D}{W}\right)^{13/3} \approx [135\,\text{ms}] + [249\,\text{ms/bit}]\log_2\left(\frac{D}{W}+1\right). \quad (2.38)$$

Performing the OLS regression over the range $1 \leq D/W \leq 64$, we find:

$$\begin{aligned}\tau &\approx 440\,\text{ms}, \\ |\phi_{11}|\tau^6 &\approx 1.4.\end{aligned} \quad (2.39)$$

The corresponding lower bound on the total movement time is $T_{LB} = 350$ ms. We provide a comparison of the functional form of the model in Eq. (2.35) to Eq. (2.3) for this case in Figure 2.4. We note that the approximations used to derive Eq. (2.35) break down for $D/W$ greater than about 50.

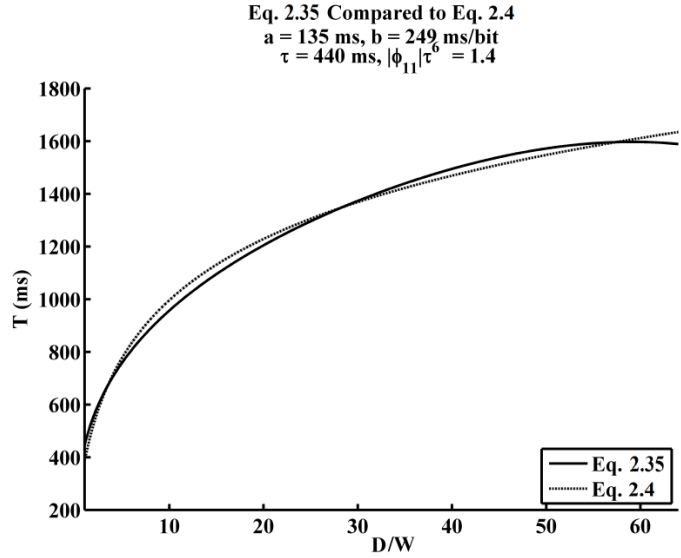

**Figure 2.4.** Eq. (2.35) compared to Eq. (2.4). The model in Eq. (2.35) is compared to the empirical formulation of Fitts' law in Eq. (2.4) for the case of $a = 135$ ms and $b = 249$ ms/bit over the range $1 \leq D/W \leq 64$.

*2.4.4 Case 3: $\phi_{11} = 0$*

We finally consider the case of $\phi_{11} = 0$. We place the same restrictions on $\phi_{33}$ as were placed on $\phi_{11}$ in Section 2.4.2, allowing it to be an unknown function of the distance to target $D$ and the target width $W$, but not of the movement time $T$ or the generalized energy $\Psi$.

In the case of $\phi_{11} = 0$, we can rewrite Eq. (2.20) in the form:

$$\tau^{-6}\left(\frac{W}{D}\right)^2 T^6 - \frac{34}{75}\phi_{33}{}^2 T^4 + \frac{4}{5}\phi_{33}T^2 - 1 \approx 0. \qquad (2.40)$$

We note that Eq. (2.40) is cubic in $T^2$. An approximate solution to Eq. (2.40) that is suitable for estimating the value of $\phi_{33}$ given values of $D/W$ and $T$ is (see Section 2.7.3):

$$T \approx \tau \cdot \left(\frac{D}{W}\right)^{1/3} \left(1 - \frac{2\phi_{33}\tau^2}{15}\left(\frac{D}{W}\right)^{2/3}\right). \tag{2.41}$$

We can look at Eq. (2.41) as being the sum of Eq. (2.29) and a corrective term of first-order in $\phi_{33}$.

In the case of point-select movements, where the subject *does not* hold down the mouse button during the movement, using the Fitts' law from Eq. (2.3), we find:

$$\tau \cdot \left(\frac{D}{W}\right)^{1/3} - \frac{2\tau \cdot \left(\phi_{33}\tau^2\right)}{15} \cdot \left(\frac{D}{W}\right) \approx [230\,\text{ms}] + [166\,\text{ms/bit}]\log_2\left(\frac{D}{W}+1\right). \tag{2.42}$$

Performing the OLS regression over the range $1 \leq D/W \leq 64$, we find:

$$\begin{aligned} \tau &\approx 400\,\text{ms}, \\ \phi_{33}\tau^2 &\approx 0.11. \end{aligned} \tag{2.43}$$

The corresponding lower bound on the total movement time is $T_{LB} = 310$ ms. We provide a comparison of the functional form of the model in Eq. (2.41) to Eq. (2.3) for this case in Figure 2.5.

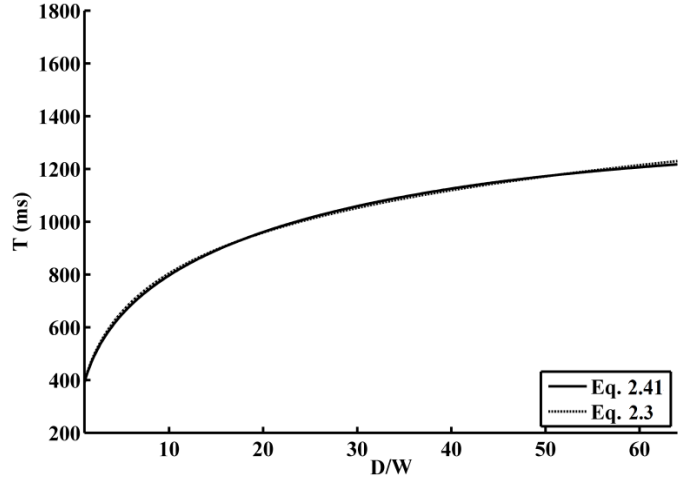

**Figure 2.5.** Eq. (2.41) compared to Eq. (2.3). The model in Eq. (2.41) is compared to the empirical formulation of Fitts' law in Eq. (2.3) for the case of $a = 230$ ms and $b = 166$ ms/bit over the range $1 \leq D/W \leq 64$.

In the case of drag-select and stroke-through movements, where the subject *does* hold down the mouse button during the movement, using the Fitts' law from Eq. (2.4), we find:

$$\tau \cdot \left(\frac{D}{W}\right)^{1/3} - \frac{2\tau \cdot \left(\phi_{33}\tau^2\right)}{15} \cdot \left(\frac{D}{W}\right) \approx [135\,\text{ms}] + [249\,\text{ms/bit}]\log_2\left(\frac{D}{W}+1\right). \qquad (2.44)$$

Performing the OLS regression over the range $1 \leq D/W \leq 64$, we find:

$$\begin{aligned} \tau &\approx 480\,\text{ms}, \\ \phi_{33}\tau^2 &\approx 0.07. \end{aligned} \qquad (2.45)$$

The corresponding lower bound on the total movement time is $T_{LB} = 380$ ms. We provide a comparison of the functional form of the model in Eq. (2.41) to Eq. (2.4) for this case in Figure 2.6.

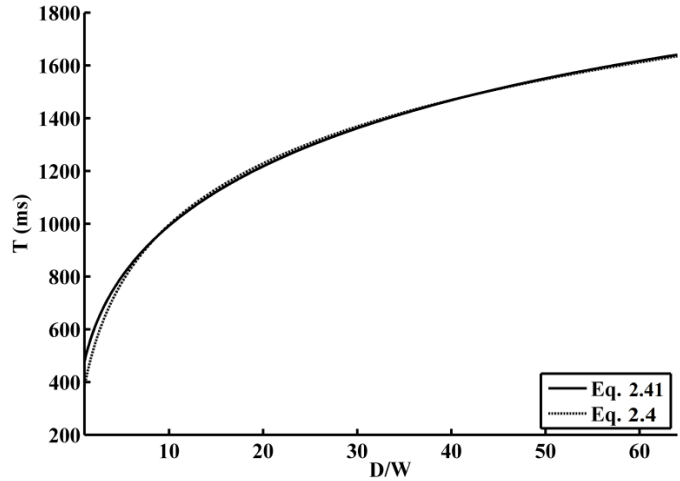

**Figure 2.6.** Eq. (2.41) compared to Eq. (2.4). The model in Eq. (2.41) is compared to the empirical formulation of Fitts' law in Eq. (2.4) for the case of $a$ = 135 ms and $b$ = 249 ms/bit over the range $1 \leq D/W \leq 64$.

*2.4.5 Remarks*

In the model in developed Section 2.3, the parameter $\phi_{11}$ describes the rate of the accrual of the cost that is associated with the mouse being at various positions relative to the target region. We have required that the rate of the accrual of the cost be approximately constant within the target region. Thus $\phi_{11}$ represents a kind of weighting the subject puts on having the mouse linger at various positions relative to the target region. The parameter $\phi_{33}$, on the other hand, describes the rate of the accrual of the cost that is associated with the acceleration of the mouse. We can interpret the parameter $\phi_{33}$ associated with higher accruals of the cost to be telling us that the subject finds it harder to accelerate the mouse, while values associated with lower accruals of the cost tell us that the subject finds it easier to accelerate the mouse. As the parameter $\phi_{33}$ appears with a minus sign in Eq. (2.10), we find that lower values of $\phi_{33}$ are associated with a higher accrual of cost, and higher values with a lower accrual of the cost.

We have obtained the best approximations to Fitts' law using the model in Section 2.3.4 where $\phi_{11} = 0$, and $\phi_{33}$ is used to fit the data. Thus, the time expressed to make the mouse movement according to Fitts' law appears to depend on how easy the subject finds it to accelerate the mouse, but not on any costs the subject associates with positions relative to the target. We also observe that the point-select movements in which the subject holds mouse button down are associated with a higher value of $\phi_{33}$ than the drag-select and stroke-through movements in which the subject does not hold mouse button down.

We have observed that $\phi_{11}$ does not appear to affect Fitts' law for mouse movements. However, mathematically, its presence in Eq. (2.10) *does* introduce an asymmetry into the velocity curve $\dot{x}(t)$ of the orbit over the course of the movement. Models for the rapid, targeted movements described by Fitts' law have movements with asymmetric velocity curves so that the movement begins with a quick acceleration to the peak velocity and then slows down gradually as the movement zeroes in on the target. So although it appears that, although $\phi_{11}$ is not necessary to derive an approximation to Fitts' law, it is nevertheless a necessary part of the model for rapid, targeted movements.

## 2.5 Alternative Control Models of Mouse Movements

We now look at alternative optimal control models for mouse movements built on control of the various time-derivatives of the orbit $d^n x / dt^n$ (i.e. any time-derivative of the orbit including the jerk). In each of these alternative control models, we construct an estimate of Fitts' law analogous to that given for jerk-control in Section 2.4.2. In each case, we arrive at a formulation Fitts' law by assuming that the generalized energy $\Psi$ is proportional to the square of the target width $W$ as in Eq. (2.28). We then compare the performance of the resulting estimate of Fitts' law to the empirical form of Fitts' law in Eq. (2.1) to see if the empirical form of Fitts' law provides empirical evidence for what quantity the human body controls during movement.

In general, a model using the control $d^n x / dt^n$ involves a more general $\Phi(x,\ldots,d^{n-1}x/dt^{n-1})$ that includes time-derivatives of the orbit up to on order below that of the control for the model. Thus, in general, the alterative control models have different numbers of free parameters $\phi_{mn}$. We have found that, for the optimal jerk-control model explored in Section 2.4, $\Phi(x,\dot{x},\ddot{x})$ has the effect of providing a relatively small modification to the form of Fitts' law found by setting $\Phi(x,\dot{x},\ddot{x}) = 0$, and so we assume that, in the general models, the $\Phi(x,\ldots,d^{n-1}x/dt^{n-1})$ would also provide relatively small modifications to the forms of Fitts' law found by setting $\Phi(x,\ldots,d^{n-1}x/dt^{n-1}) = 0$. Thus, we argue that the best model is the one that gives the closest approximation to Fitts' law when $\Phi(x,\ldots,d^{n-1}x/dt^{n-1}) = 0$.

The general optimal control problem for basic movements in the style of the optimal jerk-control model developed in Chapter 1 with $\Phi(x,\ldots,d^{n-1}x/dt^{n-1}) = 0$ is that of minimizing the cost functional $J[\cdot]$ under the initial and final conditions given by:

$$
\begin{aligned}
J[x] &= \int_0^T \frac{1}{2}\left(\frac{d^n x}{dt^n}\right)^2 dt, \\
x(0) &= -D, \qquad x(T) = 0, \\
&\ldots, \qquad\qquad \ldots, \\
\frac{d^{n-1}x}{dt^{n-1}}(0) &= 0, \quad \frac{d^{n-1}x}{dt^{n-1}}(T) = 0.
\end{aligned}
\qquad (2.46)
$$

We note that in Eq. (2.46), all time derivatives of the position up to the $n-1$-th are zero at the beginning and end of the movement. The formula corresponding to Eq. (2.18) has the form:

$$\frac{d^n x}{dt^n}(0) \sim \frac{D}{T^n}. \tag{2.47}$$

The formula corresponding to Eq. (2.26) has the form:

$$T^{2n} \sim \frac{D^2}{\Psi}. \tag{2.48}$$

Plugging an appropriately valued $\Psi \sim W^2 / \tau^n$ corresponding to Eq. (2.28) into Eq. (2.48) gives the estimated form of Fitts' law:

$$T = \tau \cdot \left(\frac{D}{W}\right)^{1/n}. \tag{2.49}$$

Eq. (2.49) generalizes the optimal jerk-control in Eq. (2.29). For general $n$, Eq. (2.49) corresponds to special cases of Kvålseth's law for rapid, targeted movements. [38] For $n = 1$, Eq. (2.49) corresponds to the estimated movement time for the free classical mechanical particle in Eq. (2.25) where the classical energy is chosen according to $E/m \sim W^2$. For $n = 2$, Eq. (2.49) gives a form of Meyer's law, a stochastic optimized-submovement model of rapid, targeted movements. [39] For $n = 3$, Eq. (2.49) gives the estimate of Fitts' law using jerk-control as calculated in Eq. (2.29).

In the case of point-select movements, where the subject *does not* hold down the mouse button during the movement, using the Fitts' law from Eq. (2.3), we find:

$$\tau \cdot \left(\frac{D}{W}\right)^{1/n} \approx [230\,\text{ms}] + [166\,\text{ms/bit}] \log_2\left(\frac{D}{W}+1\right). \tag{2.50}$$

In Figure 2.7, we fit several models to Eq. (2.3) using OLS regression over the range $1 \leq D/W \leq 64$ and compare the resulting functional forms.

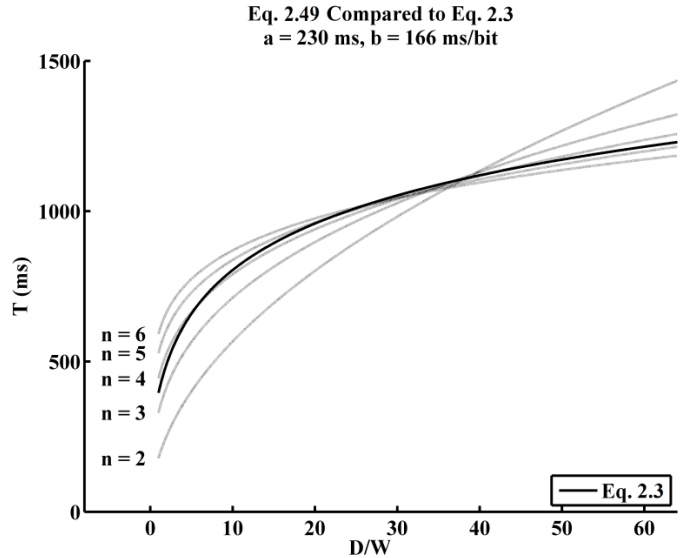

**Figure 2.7.** Eq. (2.49), compared to Eq. (2.3). The alternate control models specified by different values of $n$ in Eq. (2.49) are compared to the empirical formulation of Fitts' law in Eq. (2.3) for the case of $a$ = 230 ms and $b$ = 166 ms/bit over the range $1 \leq D/W \leq 64$.

In the case of drag-select and stroke-through movements, where the subject *does* hold down the mouse button during the movement, using the Fitts' law from Eq. (2.4), we find:

$$\tau \cdot \left(\frac{D}{W}\right)^{1/n} \approx [135\,\text{ms}] + [249\,\text{ms/bit}] \log_2\left(\frac{D}{W}+1\right). \tag{2.51}$$

In Figure 2.8, we fit several models to Eq. (2.4) using OLS regression over the range $1 \leq D/W \leq 64$ and compare the resulting functional forms.

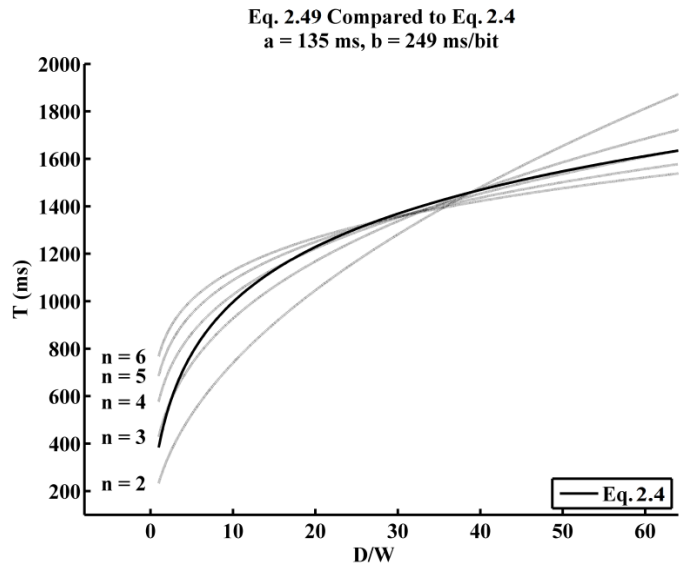

**Figure 2.8.** Eq. (2.49), compared to Eq. (2.4). The alternate control models specified by different values of $n$ in Eq. (2.49) are compared to the empirical formulation of Fitts' law in Eq. (2.4) for the case of $a$ = 135 ms and $b$ = 249 ms/bit over the range $1 \leq D/W \leq 64$.

In Chapter 1, we developed the formalism of jerk-control by arguing that we could rule out velocity- and acceleration-control due to the observation that these controls introduces discontinuities in the forces exerted by muscles when making the movements. We argued that muscles could not jump discontinuously from exerting one force to exerting another without exerting all the intermediate forces during some transition interval. In Figure 2.7, the empirical form of Fitts' law in Eq. (2.3) lies near to forms of Eq. (2.49) with values $n$ of 3, 4, or 5 (i.e. jerk-, snap-, or crackle-control). In Figure 2.8, the empirical form of Fitts' law in Eq. (2.4) lies near to forms of Eq. (2.49) with values $n$ of 2, 3, or 4 (i.e. acceleration-, jerk-, or snap-control). Given the argument in Chapter 1 together, Figures 2.7 and 2.8 suggest that the body uses either jerk- or snap-control.

## 2.6 Discussion

We have used the general model for optimal jerk-control of basic human movements to construct a model that describes movements of a computer mouse using the hand. We have shown that we can introduce Fitts' law into the mathematical model of motor behavior developed in Chapter 1 by supposing that, for rapid, targeted movements, the subject making the movement chooses the generalized energy $\Psi$ to be proportional to the square of the width $W$ of the target region. We have finally shown that, given the constraints of the optimal control movement problem, choice of the generalized energy $\Psi$ determines the total movement time $T$.

Fitts' law describes the speed/accuracy trade off in rapid, targeted movements. We should understand the subject's choice of the generalized energy $\Psi$ for a movement as reflecting how often the subject is willing to miss the target. Since the value of the generalized energy is determined by the value of the parameter $\tau$ in Eq. (2.28), we expect $\tau$ is determined by the choice of how often the subject is willing to miss the target. In a more complete model, we would expect there to be a formula relating $\tau$ to the acceptable error-rate.

Given the acceptable error-rate, the total movement time $T$ reflects the time of the quickest movement that the subject can make given the distance $D$ to the target and the width $W$ of the target region. The subject can always choose to make slower movements (probably giving an actual error-rate below the acceptable error-rate) having longer total movement times. Thus, the total movement time $T$ given by Fitts' law reflects a lower bound on the movement time. As a result, the generalized energy $\Psi$ associated with the total movement time given by Fitts' law reflects an upper bound on the generalized energy.

## 2.7 Appendices

### 2.7.1 Appendix 1

We would like to find an approximation for the orbit $x(t)$ of a computer mouse movement that begins motionless at a position $-D$ and ends motionless at the origin at some time $T$. Formally, we express these conditions as:

$$\begin{aligned} x(0) &= -D, & x(T) &= 0, \\ \dot{x}(0) &= 0, & \dot{x}(T) &= 0, \\ \ddot{x}(0) &= 0, & \ddot{x}(T) &= 0. \end{aligned} \qquad (2.52)$$

The equation of motion of the system follows from Eq. (2.12), that is:

$$\ddddot{x} + \phi_{33}\dddot{x} - (\phi_{22} - \phi_{13})\ddot{x} + \phi_{11}x = 0. \qquad (2.53)$$

Although Eq. (2.53) can be solved exactly using standard mathematical technique, it is more convenient to generate an approximate solution to Eq. (2.12) in the form of a truncated Taylor series expansion about time $t = 0$:

$$\begin{aligned} x(t) &\approx x(0) + \dot{x}(0)t + \frac{1}{2}\ddot{x}(0)t^2 + \frac{1}{6}\dddot{x}(0)t^3 \\ &+ \frac{1}{24}\ddddot{x}(0)t^4 + \frac{1}{120}\dddddot{x}(0)t^5 + \frac{1}{720}\ddddddot{x}(0)t^6. \end{aligned} \qquad (2.54)$$

The unknown coefficients on the RHS of Eq. (2.54) must satisfy Eqs. (2.52) and (2.53), and as there are seven unknown coefficients in Eq. (2.54) and seven equations in Eqs. (2.52) and (2.53), there is a unique solution. Combining Eqs. (2.52) and (2.54) using the conditions for time $t = 0$, and evaluating Eq. (2.53) at time $t = 0$ gives the following system of equations:

$$\begin{aligned} x(t) &\approx -D + \frac{1}{6}\dddot{x}(0)t^3 + \frac{1}{24}\ddddot{x}(0)t^4 + \frac{1}{120}\dddddot{x}(0)t^5 + \frac{1}{720}\ddddddot{x}(0)t^6, \\ \dot{x}(t) &\approx \frac{1}{2}\dddot{x}(0)t^2 + \frac{1}{6}\ddddot{x}(0)t^3 + \frac{1}{24}\dddddot{x}(0)t^4 + \frac{1}{120}\ddddddot{x}(0)t^5, \\ \ddot{x}(t) &\approx \dddot{x}(0)t + \frac{1}{2}\ddddot{x}(0)t^2 + \frac{1}{6}\dddddot{x}(0)t^3 + \frac{1}{24}\ddddddot{x}(0)t^4, \\ \ddddot{x}(0) &+ \phi_{33}\dddot{x}(0) = \phi_{11}D. \end{aligned} \qquad (2.55)$$

Evaluating Eq. (2.55) at $t = T$ gives:

$$D \approx \frac{1}{6}\dddot{x}(0)T^3 + \frac{1}{24}\ddddot{x}(0)T^4 + \frac{1}{120}\dddddot{x}(0)T^5 + \frac{1}{720}\ddddddot{x}(0)T^6,$$

$$0 \approx \frac{1}{2}\dddot{x}(0)T^2 + \frac{1}{6}\ddddot{x}(0)T^3 + \frac{1}{24}\dddddot{x}(0)T^4 + \frac{1}{120}\ddddddot{x}(0)T^5,$$

$$0 \approx \dddot{x}(0)T + \frac{1}{2}\ddddot{x}(0)T^2 + \frac{1}{6}\dddddot{x}(0)T^3 + \frac{1}{24}\ddddddot{x}(0)T^4, \quad (2.56)$$

$$\dddddot{x}(0) + \phi_{33}\dddot{x}(0) = \phi_{11}D.$$

We can rewrite the system of equations in Eq. (2.56) compactly in matrix form as:

$$\begin{bmatrix} 720D \\ 0 \\ 0 \\ (\phi_{11}T^6)D \end{bmatrix} \approx \begin{bmatrix} 120 & 30 & 6 & 1 \\ 60 & 20 & 5 & 1 \\ 24 & 12 & 4 & 1 \\ 0 & (\phi_{33}T^2) & 0 & 1 \end{bmatrix} \begin{bmatrix} \dddot{x}(0)T^3 \\ \ddddot{x}(0)T^4 \\ \dddddot{x}(0)T^5 \\ \ddddddot{x}(0)T^6 \end{bmatrix}. \quad (2.57)$$

We note that the values $\phi_{11}T^6$ and $\phi_{33}T^2$ are unitless, that is they are just real numbers. We can simplify Eq. (2.57) using standard row operations. First, we use the third equation to eliminate terms in $\ddddddot{x}(0)$ from the first and second equations:

$$\begin{bmatrix} 2880D \\ 0 \\ 0 \\ (\phi_{11}T^6)D \end{bmatrix} \approx \begin{bmatrix} 336 & 48 & 0 & -2 \\ 120 & 20 & 0 & -1 \\ 24 & 12 & 4 & 1 \\ 0 & (\phi_{33}T^2) & 0 & 1 \end{bmatrix} \begin{bmatrix} \dddot{x}(0)T^3 \\ \ddddot{x}(0)T^4 \\ \dddddot{x}(0)T^5 \\ \ddddddot{x}(0)T^6 \end{bmatrix}. \quad (2.58)$$

Next, we use the fourth equation to eliminate terms in $\ddddddot{x}(0)$ from the remaining equations:

$$\begin{bmatrix} (2880 + 2(\phi_{11}T^6))D \\ (\phi_{11}T^6)D \\ -(\phi_{11}T^6)D \\ (\phi_{11}T^6)D \end{bmatrix} \approx \begin{bmatrix} 336 & 48+2(\phi_{33}T^2) & 0 & 0 \\ 120 & 20+(\phi_{33}T^2) & 0 & 0 \\ 24 & 12-(\phi_{33}T^2) & 4 & 0 \\ 0 & (\phi_{33}T^2) & 0 & 1 \end{bmatrix} \begin{bmatrix} \dddot{x}(0)T^3 \\ \ddddot{x}(0)T^4 \\ \dddddot{x}(0)T^5 \\ \ddddddot{x}(0)T^6 \end{bmatrix}. \quad (2.59)$$

Finally, we use the second equation to eliminate terms in $\ddddot{x}(0)$ from the first equation leaving:

$$(336(20+(\phi_{33}T^2))-120(48+2(\phi_{33}T^2)))\ddot{x}(0)T^3 \approx ((20+(\phi_{33}T^2))(2880+2(\phi_{11}T^6))-(48+2(\phi_{33}T^2))(\phi_{11}T^6))D. \tag{2.60}$$

This gives:

$$\ddot{x}(0) \approx \left(\frac{7200+360(\phi_{33}T^2)-(\phi_{11}T^6)}{120+54(\phi_{33}T^2)}\right)\frac{D}{T^3}. \tag{2.61}$$

The remaining terms $\dddot{x}(0)$, $\ddddot{x}(0)$, and $\dddddot{x}(0)$ may be evaluated in a similar manner. If we assume that $\phi_{11}T^6$ and $\phi_{33}T^2$ is relatively small, then we can approximate Eq. (2.61) using a Taylor series expansion. Keeping to at most second order in $\phi_{11}T^6$ and $\phi_{33}T^2$, we find that Eq. (2.61) is approximately:

$$\ddot{x}(0) \approx \left(60 - 24\phi_{33}T^2 - \frac{\phi_{11}T^6}{120} + \frac{54\phi_{33}^2T^4}{5} + \frac{9\phi_{11}\phi_{33}T^8}{20}\right)\frac{D}{T^3}. \tag{2.62}$$

*2.7.2 Appendix 2*

We would like estimate the value of $\phi_{11}$ given values of $D/W$ and $T$ which satisfy the quadratic equation in $T^6$ given by:

$$\frac{\phi_{11}^2}{4\cdot 60^4}T^{12} - \tau^{-6}\left(\frac{W}{D}\right)^2 T^6 + 1 \approx 0. \tag{2.63}$$

We proceed to derive an approximate solution to Eq. (2.63) that is suitable for fitting to data to provide an estimate of $\phi_{11}$. We can evaluate Eq. (2.63) using the quadratic formula. Limiting to solutions of Eq. (2.34) that give physically sensible movement times, we find:

$$T^6 \approx \frac{2\cdot 60^4}{\phi_{11}^2\tau^6}\left(\frac{W}{D}\right)^2\left(1 - \sqrt{1 - \frac{(\phi_{11}\tau^6)^2}{60^4}\left(\frac{D}{W}\right)^4}\right). \tag{2.64}$$

Taking the Taylor series expansion of the square root in Eq. (2.64) and truncating the expansion to second order in $\phi_{11}^2$, we find:

$$T^6 \approx \frac{2 \cdot 60^4}{\phi_{11}^2 \tau^6}\left(\frac{W}{D}\right)^2 \left[\frac{(\phi_{11}\tau^6)^2}{2 \cdot 60^4}\left(\frac{D}{W}\right)^4 - \frac{(\phi_{11}\tau^6)^4}{8 \cdot 60^8}\left(\frac{D}{W}\right)^8\right]. \tag{2.65}$$

We can simplify Eq. (2.65) to the form:

$$T^6 \approx \tau^6 \left(\frac{D}{W}\right)^2 \left[1 - \frac{(\phi_{11}\tau^6)^2}{4 \cdot 60^4}\left(\frac{D}{W}\right)^4\right]. \tag{2.66}$$

Taking the sixth root and expanding the RHS to first order this gives the first order correction:

$$T \approx \tau \cdot \left(\frac{D}{W}\right)^{1/3} \left[1 - \frac{(\phi_{11}\tau^6)^2}{24 \cdot 60^4}\left(\frac{D}{W}\right)^4\right]. \tag{2.67}$$

*2.7.3 Appendix 3*

We would like estimate the value of $\phi_{33}$ given values of $D/W$ and $T$ which satisfy the cubic equation in $T^2$ given by:

$$\tau^{-6} \cdot \left(\frac{W}{D}\right)^2 T^6 - \frac{34}{75}\phi_{33}^2 T^4 + \frac{4}{5}\phi_{33} T^2 - 1 \approx 0. \tag{2.68}$$

We proceed to derive an approximate solution to Eq. (2.68) that is suitable for fitting to data to provide an estimate of $\phi_{33}$. While we could evaluation Eq. (2.68) using the cubic formula, it is easier to approximate the solution using perturbation theory (see e.g., [40]). We begin by assuming that $T^2$ approximately has the form:

$$T^2 \approx \Theta_0 + \phi_{33}\Theta_1. \tag{2.69}$$

The square and cube of Eq. (2.69) are:

$$\begin{aligned}T^4 &\approx \Theta_0^2 + 2\phi_{33}\Theta_0\Theta_1 + \phi_{33}^2\Theta_1^2, \\ T^6 &\approx \Theta_0^3 + 3\phi_{33}\Theta_0^2\Theta_1 + 3\phi_{33}^2\Theta_0\Theta_1^2 + \phi_{33}^3\Theta_1^3.\end{aligned} \tag{2.70}$$

Since we only expand $T^2$ to first-order in $\phi_{33}$ we only keep to first-order in $\phi_{33}$ when plugging the approximation into Eq. (2.68). We find:

$$\left(\tau^{-6} \cdot \left(\frac{W}{D}\right)^2 \Theta_0^3 - 1\right) + \phi_{33} \cdot \left(\frac{4}{5}\Theta_0 + 3\tau^{-6} \cdot \left(\frac{W}{D}\right)^2 \Theta_0^2 \Theta_1\right) \approx 0. \qquad (2.71)$$

We evaluate $\Theta_0$ and $\Theta_1$ by setting the two terms on the LHS of Eq. (2.71) to zero separately:

$$\Theta_0 \approx \tau^2 \cdot \left(\frac{D}{W}\right)^{2/3}, \quad \Theta_1 \approx -\frac{4\tau^4}{15}\left(\frac{D}{W}\right)^{4/3}. \qquad (2.72)$$

Plugging Eq. (2.72) into Eq. (2.71) we find:

$$T^2 \approx \tau^2 \cdot \left(\frac{D}{W}\right)^{2/3}\left(1 - \frac{4\phi_{33}\tau^2}{15}\left(\frac{D}{W}\right)^{2/3}\right). \qquad (2.73)$$

Taking the square root of both sides of Eq. (2.73), and approximating the square root on the RHS by truncating the Taylor series expansion of the square root to first order in $\phi_{33}$, we find:

$$T \approx \tau \cdot \left(\frac{D}{W}\right)^{1/3}\left(1 - \frac{2\phi_{33}\tau^2}{15}\left(\frac{D}{W}\right)^{2/3}\right). \qquad (2.74)$$

# Chapter 3 – On the Description of Finger-Tapping

**Abstract.** We use the optimal jerk-control model of human movements developed in Chapter 1 to formulate a mathematical model describing how a subject makes a complex movement that is composed of a sequence of simple movements. In particular, we look at a relatively simple case of such a complex movement made up of simple movements – finger-tapping – that also forms the basis of a standard clinical test of motor speed, the Finger-Tapping Test. We show that we can model the complex movement by defining a sequence of intermediate states that happen between the initial and final states of the complex movements and use the optimal jerk-control model to find the movement that takes place between two consecutive states in this sequence. We look at summary measures for the performance of the complex movement derived from the generalized energy and compare them to a measure that has been proposed to characterize finger-tapping.

## 3.1 Introduction

In Chapters 1 and 2, we have developed an optimal jerk-control model and used to describe simple movements, that is, movements that minimize a cost functional while satisfying a set of initial and final conditions. While many everyday movements that people make are simple movements (e.g. the computer mouse movements we have examined in Chapter 2), many other movements are more complicated than this, and cannot be analyzed using a model for simple movements. In this Chapter, we look at one way to build up complicated movements out of a sequence of simple movements. We look specifically at a simple case of a complicated

movement – finger-tapping. Finger-tapping is itself of clinical interest as it forms the basis of a standard clinical test – the Finger-Tapping Test − a commonly used clinical test of motor speed in which the standard measure of performance is the average number of taps a subject can make in 10 s. [41, 42]

We can look at the basic movement of finger-tapping as bearing a close relationship to typing on a keyboard, and, indeed the use of keyboard typing as an in-home proxy measure for finger-tapping speed has been proposed. [9] We expect that a model of finger-tapping should inform the construction of a model of typing and indicate how to relate in-home measures of typing speed to clinical measures of finger-tapping speed.

To develop the desired framework for understanding finger-tapping using the optimal jerk-control model, we have to look at how a subject makes a continuous sequence of movements and how to make the model best describe that process. We think of finger-tapping as a single complex movement made up of a sequence of simple movements that proceed from one to the next without any pause in between. Thus, the state of the part of the body being moved during finger-tapping (the fingertip) at the end of one movement becomes the state at the beginning of the next movement. Due to the periodic nature of the finger-tapping movement the subject, has some freedom to choose to make each of the simple movements in such a way as to render the whole sequence of simple movements that make up finger-tapping easier to carry out. We observe that there is a natural way to look at how a subject may choose to make simple movements that make the complex movement easier within the framework of the model of movements developed in Chapter 1. As the original model of movement that we have developed in Chapter 1 was built on the idea of minimizing a cost functional, the subject can choose to make the simple movements in a way that minimizes the cost functional for the entire complex movement of the finger-tapping sequence. We expect that the framework developed for looking at finger-tapping can be generalized to other kinds of periodic human movements.

When looking at complex movements made up of a sequence of simple movements, it is useful to find a way to characterize the entire complex movement in terms of sum set of summary values. One would like these values to sum up the entire complex movement rather than each of the simple movements making it up. We have observed in Chapters 1 and 2 that the generalized energy is a value that is constant for each simple movement and provides information about how "hard," or "strong," or "fast" that simple movement is. It is natural, then, to look at the average generalized energy for an entire complex movement as a measure of how "hard," or "strong," or "fast" that complex movement is. We look at this value for finger-tapping and how it relates to a finger-tapping performance metric proposed in [43].

**3.2 Complex Movements**

We approach the analysis of finger-tapping by treating the whole sequence of finger-taps as a single, continuous movement composed of a sequence of simple movements like those described in Chapter 1. We call this kind of movement a *complex movement*.

In Chapters 1 and 2, we have looked at simple movements and, in particular, the special case of simple movements in which the movement begins and ends at rest that we have called a *basic movement*. Simple movements are movements that are concerned with going from the initial conditions to the final conditions in the most straightforward way that minimizes the cost. In general, humans often make much more complicated movements than can be accounted for using simple movements. In these, more complicated, movements, one must meet a sequence of intermediate conditions in the course of completing the movement. We now look at how we can expand the model developed in Chapter 1 to describe movements that are more complicated than simple movements.

We assume that we can describe a movement but specifying the orbit $x(t)$ in three-dimensional space for some point on the body. We assume that the body is constrained in such a way that the specification of the orbit of one point determines the movement of the entire body. In the case of yet more complicated movements, orbits can be specified for several points on the body and those orbits together with constraints on the body can determine the motion of the body. In the model of human movements presented here, we assume that the functions of position, velocity, and acceleration associated with the orbit $x(t)$ that describes the movement must be continuous throughout the movement and over some time immediately before and after the movement. This follows from basic physical laws and the physical assumption that we made in Chapter 1 that muscles cannot discontinuously change the forces they are applying.

We define a *simple movement* to be the orbit $x(t)$ that minimizes the cost functional given a set of initial and final conditions. Mathematically, we express the problem of finding the orbit for a simple movement as:

$$J[x] = \int_0^T L(x, \dot{x}, \ddot{x}, \dddot{x}) dt,$$
$$x(0) = x_0, \quad x(T) = x_1,$$
$$\dot{x}(0) = v_0, \quad \dot{x}(T) = v_1, \quad (3.1)$$
$$\ddot{x}(0) = a_0, \quad \ddot{x}(T) = a_1.$$

We assume that the initial and final velocities and accelerations are continuous with the velocities and accelerations before and after the movement, that is the velocity and acceleration are continuous from a point before the movement begins to a point after the movement ends. The jerk, however, may be discontinuous at the beginning and ending of the movement. The generalized energy Ψ has a constant value during the movement, though this value may differ from the value of the generalized energy before or after the movement.

In Chapter 1, in order to keep the mathematics straightforward, we have assumed that simple movements travel in one dimension along the line in space that connects the initial and final positions; an assumption that works fine in the case of the basic movements that we have used to describe mouse movements in Chapter 2. However, in general a simple movement takes place in three-dimensional space, and may move away from the line connecting the initial and final positions.

We define a *complex movement* to be a single movement that is composed out of a sequence of simple movements. We think of human movements as periods of continuous movement between periods of rest; so, we think of an individual complex movement as beginning and ending at rest. Thus, a complex movement is a more complicated analog of a basic movement that is composed of a sequence of simple movements. Alternatively, a basic movement is the trivial case of a complex movement where the complex movement consists of a single simple movement.

The first simple movement in the sequence of simple movements that comprise a complex movement begins in the initial condition for the whole movement, and the last simple movement in the sequence ends in the final condition for the whole movement. Otherwise, all the simple movements in the sequence begin in the condition in which the previous simple movement has ended. We assume that the initial and final conditions for each simple movement making up the complex movement are given. Thus, the problem of finding the desired complex movement is that of solving the set of simple movement problems given by:

$$\begin{aligned}
J[x] &= \int_{T_n}^{T_{n+1}} L(x, \dot{x}, \ddot{x}, \dddot{x}) \, dt, \\
x(T_n) &= x_n, \quad x(T_{n+1}) = x_{n+1}, \\
\dot{x}(T_n) &= v_n, \quad \dot{x}(T_{n+1}) = v_{n+1}, \\
\ddot{x}(T_n) &= a_n, \quad \ddot{x}(T_{n+1}) = a_{n+1}.
\end{aligned} \quad (3.2)$$

We require that the functions of position, velocity, and acceleration of the resulting orbit $x(t)$ for the complex movement be continuous throughout the movement, but that the jerk may have discontinuities where one simple movement ends and another simple movement begins. We also assume that the initial and final velocities and accelerations are continuous with the velocities and accelerations before and after the movement, that is the velocity and acceleration are continuous from a point before the movement begins to a point after the movement ends. While the generalized energy $\Psi$ is constant for each simple movement making up the complex movement, it may not be constant over the entire complex movement. A complex movement generally takes place in three-dimensional space. The requirement that the functions of position, velocity, and acceleration be continuous means that the component simple movements will generally not travel along the lines connecting the initial and final positions.

### 3.3 Finger-Tapping

Finger-tapping is a single continuous sequence of upward and downward movements of the fingertip that lasts for a defined period. Mathematically, we treat finger-tapping as a single complex movement that begins and ends with the finger at rest and consists of a sequence of simple movements. The simple movements are the alternating upward and downward movements of the fingertip. There are a number of ways to generate a complex movement using a sequence of simple upward and downward movements. Without empirical information about how a subject performs finger-tapping, we do not know which model best describes finger-tapping in general, or, indeed, whether different subjects tap in a ways best described by different models. Moreover, as we are using finger-tapping, in part, as an application that we can use to illustrate how to approach complex movements, by examining several models of finger-tapping we can illustrate several ways of modelling complex movements. We analyze finger-tapping using two broad types of models: (1) uniform finger-tapping models, and (2) non-uniform finger-tapping models.

In the uniform finger-tapping models, we assume that every upward simple movement of the fingertip during finger-tapping is identical and that every downward simple movement of the fingertip is identical. We can think of uniform finger-tapping as having a constant amplitude and frequency for the duration of finger-tapping.

In the non-uniform finger-tapping models, we allow every simple movement of the fingertip that happens during finger-tapping to differ from the others. Thus, the amplitude and timing of each tap of the fingertip can differ from tap to tap. We can think of non-uniform finger-tapping as having only an average amplitude and frequency for the duration of finger-tapping.

For finger-tapping, there is, in practice, an obstruction at extreme downward position. This obstruction can be the surface of the table on which the hand is resting, or a lever arm that is part of a device to count finger-taps. Moreover, the finger also has limited range of motion. To keep the analysis straightforward for the models of both uniform and non-uniform finger-tapping, we ignore any consideration of how the fingertip is constrained to stop at either extreme of the tapping movement. We assume the subject just stops the finger by design at the given positions. We further assume that we can treat the motion of the fingertip during finger-tapping as always lying approximately in a line connecting the extreme upward and extreme downward positions of the fingertip. Finally, we ignore the first and last movements in the finger-tapping sequence (i.e. those that must begin and end at rest), and treat only the body of the finger-tapping movement;

these movements can, of course, be solved for once the main body of the finger-tapping movement has been described.

### 3.4 Uniform Finger-Tapping Models

We begin by looking at models of uniform finger-tapping. In uniform finger-tapping, it is assumed that the all the upward simple movements of the fingertip are assumed to be identical and all the downward simple movements of the fingertip are assumed to be identical. We define the total distance between the extreme upward and extreme downward positions of the fingertip to be a distance $2A$, where $A$ is the amplitude of the movement. We define the time taken for the fingertip to move from and return to the extreme downward position to be the period $T$ of the finger-tapping movement.

*3.4.1 Sinusoidal Finger-Tapping*

One immediate way of looking at finger-tapping is to look at the motion of the fingertip as being described by a simple harmonic oscillator. In this model, the motion of the fingertip is sinusoidal:

$$x(t) = -A\cos\left(\frac{2\pi t}{T}\right). \tag{3.3}$$

In Eq. (3.3), we have assumed that tapping beings at time $t = 0$, and that at this time the fingertip is at its extreme downward position. The movement of the fingertip in Eq. (3.3) satisfies the sequence of intermediate conditions (where $n$ is a non-negative integer):

$$\begin{aligned} x(nT) &= -A, & x\left(\left(n+\tfrac{1}{2}\right)T\right) &= A, \\ \dot{x}(nT) &= 0, & \dot{x}\left(\left(n+\tfrac{1}{2}\right)T\right) &= 0, \\ \ddot{x}(nT) &= \frac{4\pi^2 A}{T^2}, & \ddot{x}\left(\left(n+\tfrac{1}{2}\right)T\right) &= -\frac{4\pi^2 A}{T^2}, \end{aligned} \tag{3.4}$$

We observe that, in this model, when the fingertip is at the extreme positions of its motion, the velocity of the fingertip is zero, as we would expect. However, when the fingertip is at these positions, the acceleration is already working to begin movement of the fingertip to back to the other extreme position. Thus, in this model, when the subject's fingertip reaches the extreme position, the subject is already using the muscles to apply force to the fingertip.

We can solve for the jerk over the course of the finger-tapping movement; it is:

$$\dddot{x}(t) = \frac{8\pi^3 A}{T^3} \sin\left(\frac{2\pi t}{T}\right). \tag{3.5}$$

We observe that, in this model, the jerk is continuous during finger-tapping, as are all of the higher-order time-derivatives of the orbit $x(t)$. The sinusoidal model describes finger-tapping using a simple harmonic oscillator. A natural way to characterize the motion of a simple harmonic oscillator in a single value is by using the classical energy associated with the motion. The classical energy is conserved throughout the motion of a simple harmonic oscillator and provides a single value that describes the motion at all times; this value is:

$$E = \frac{2\pi^2 A^2}{T^2}. \tag{3.6}$$

It is possible to arrive at this sinusoidal finger-tapping model by minimizing a control functional using a set of initial and final conditions related to Eq. (3.4) for each simple movement in the finger-tapping sequence. However, it is not possible to do using the jerk-control model developed in Chapter 1. It is better, from our perspective, to think of the sinusoidal finger-tapping model as an approximation that is useful in certain circumstances.

### 3.4.2 Uniform Basic Movement Finger-Tapping

We can think of the sinusoidal finger-tapping model as describing finger-tapping as a sequence of simple movements where each simple movement is a segment of sinusoid comprising half of a single period. We may construct a similar model using the optimal jerk-control model that we have developed in Chapter 1 by replacing the simple movements in the sinusoidal finger-tapping model with appropriate orbits that minimize the jerk-control cost functional $J[\cdot]$. One way to do this is to construct the simple movements using basic movements that begin and end at rest. We express this mathematically by finding the orbits $x(t)$ that minimize the cost functional and satisfy intermediate conditions given by:

$$J[x] = \int_{nT}^{(n+1/2)T} \left( \frac{1}{2}\dddot{x}^2 - \Phi(x,\dot{x},\ddot{x}) \right) dt,$$

$$x(nT) = -A, \quad x\left(\left(n+\frac{1}{2}\right)T\right) = A,$$

$$\dot{x}(nT) = 0, \quad \dot{x}\left(\left(n+\frac{1}{2}\right)T\right) = 0,$$

$$\ddot{x}(nT) = 0, \quad \ddot{x}\left(\left(n+\frac{1}{2}\right)T\right) = 0.$$

(3.7)

To keep the finger-tapping problem straightforward, we also assume the effect of $\Phi(x,\dot{x},\ddot{x})$ is small and take $\Phi(x,\dot{x},\ddot{x}) = 0$. The Lagrangian for each finger-tapping movement becomes:

$$L = \frac{1}{2}\dddot{x}^2.$$

(3.8)

The Euler-Lagrange equation for Eq. (3.8) is:

$$\ddot{\dddot{x}} = 0.$$

(3.9)

We can see that the orbits of all the simple movements in the finger-tapping sequence take on the same shape. Thus, it suffices to look in detail at the movement that begins at time $t = 0$; the solution for this movement is:

$$x(t) = A \cdot \left( -1 + \frac{160}{T^3}t^3 - \frac{480}{T^4}t^4 + \frac{384}{T^5}t^5 \right).$$

(3.10)

We compare finger-tapping movements of the form in Eq. (3.10) to sinusoidal finger-tapping described by Eq. (3.3) in Figure 3.1.

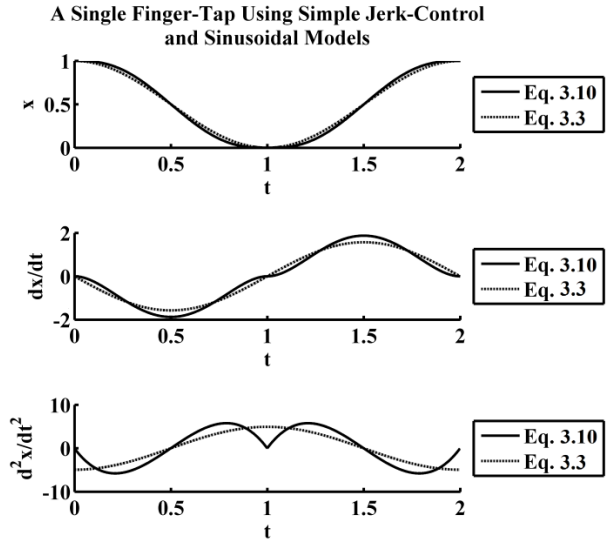

**Figure 3.1.** A single finger-tap using simple jerk-control and sinusoidal models. We compare a the form of the orbit of a single finger-tap using the jerk-control model in Eq. (3.10) to the sinusoidal model in Eq. (3.3). Note the cusps in the acceleration of the orbit ($dx^2/dt^2$) at the beginning and end of each simple movement; these correspond to discontinuities in the jerk of the orbit.

In this model, when the fingertip is at the extreme positions of its motion, the velocity of the fingertip is still zero, as we would expect. However, when the fingertip is at these positions, the acceleration is zero, so the subject is not yet using the muscles to apply forces to begin movement of the fingertip to back to the other extreme position.

We can solve for the jerk for this simple movement in the finger-tapping movement; it is:

$$\dddot{x}(t) = \frac{960A}{T^3} \cdot \left(1 - \frac{12}{T}t + \frac{24}{T^2}t^2\right). \tag{3.11}$$

The cost associated with each simple movement in the finger-tapping movement is:

$$J = \frac{920^2 A^2}{2T^3} \int_0^{T/2} \left(1 - \frac{12}{T}t + \frac{24}{T^2}t^2\right)^2 dt. \tag{3.12}$$

The generalized energy associated with each simple movement in the finger-tapping movement is:

$$\Psi = \frac{920^2 A^2}{2T^6}. \tag{3.13}$$

We have kept this model mathematically straightforward by requiring the simple movements that make up the complex movement of finger-tapping to be basic movements, that is, movements that begin and end at rest. Thus, the model introduces a very short pause in all movement at the extreme upward and downward positions of the fingertip. We can eliminate this very short pause in all movement at these points by allowing the simple movements in the finger-tapping movement to have acceleration at the extreme upward and downward positions. That is, we would allow the subject to perform finger-tapping so that at the extreme upward and downward positions, the muscles are already providing the force needed to begin the next simple movement in the finger-tapping movement. This would give a model that is more like the sinusoidal finger-tapping model.

*3.4.3 Uniform Minimum Cost Simple Movement Finger-Tapping*

For basic movement finger-tapping, we have solved the problem to find the simple movements that make up the complex movement of the finger-tapping movement that satisfy the intermediate conditions that the movement being and end at rest. While this model of finger-tapping does use the simple movements that minimize the cost functional $J[\cdot]$ given the intermediate conditions in Eq. (3.7), it does not necessarily produce the finger-tapping movement that has the lowest cost. We can possibly find a lower cost finger-tapping movement by allowing some of the intermediate conditions to be free variables. We may then find the values of these free variables that minimize the cost functional.

In uniform finger-tapping, we require that the fingertip move between clearly defined extreme upward and downward positions, and that the fingertip stop at the extreme positions. Thus, we must have the same intermediate conditions in the position and velocity that we have used in Eqs. (3.4) and (3.7). The only intermediate conditions that we are free to leave as free variables are the initial and final accelerations that were non-zero in the case of sinusoidal finger-tapping in Eq. (3.4) and were zero in the case of basic movement finger-tapping in Eq. (3.7). We define the intermediate accelerations using the free variable $a$, and denote the orbit of a simple movement of finger-tapping given a value of $a$ by $x_a(t)$. Thus, each simple movement of the finger-tapping movement must solve:

$$J[x_a] = \int_{nT}^{(n+1/2)T} \left( \frac{1}{2}\ddot{x}_a^2 - \Phi(x_a, \dot{x}_a, \ddot{x}_a) \right) dt,$$

$$x_a(nT) = -A, \quad x_a\left(\left(n+\frac{1}{2}\right)T\right) = A,$$

$$\dot{x}_a(nT) = 0, \quad \dot{x}_a\left(\left(n+\frac{1}{2}\right)T\right) = 0, \quad (3.14)$$

$$\ddot{x}_a(nT) = a, \quad \ddot{x}_a\left(\left(n+\frac{1}{2}\right)T\right) = -a.$$

To keep the finger-tapping problem simple, we again assume the effect of $\Phi(x_a, \dot{x}_a, \ddot{x}_a)$ is small and take $\Phi(x_a, \dot{x}_a, \ddot{x}_a) = 0$. The Lagrangian for each finger-tapping movement again becomes:

$$L = \frac{1}{2}\ddot{x}_a^2. \quad (3.15)$$

The Euler-Lagrange equation for Eq. (3.15) is again:

$$\dddot{\ddot{x}}_a = 0. \tag{3.16}$$

We can see that the orbits of all the simple movements in the finger-tapping sequence again take on the same shape. Thus, it suffices to look in detail at the movement that begins at time $t = 0$; the solution for this movement is:

$$x_a(t) = \sum_{v=0}^{5} c_v t^v. \tag{3.17}$$

We can evaluate the $c_v$ in Eq. (3.17) by requiring $x_a(t)$ to satisfy the intermediate conditions in Eq. (3.14), although we do not do that here.

The cost associated with each simple movement in the finger-tapping movement is:

$$J[x_a] = \frac{1}{2} \int_0^{T/2} \dddot{x}_a(t)^2 \, dt. \tag{3.18}$$

We find the value of the acceleration $a$ that minimizes the cost of finger-tapping by taking:

$$\frac{dJ[x_a]}{da} = 0. \tag{3.19}$$

The generalized energy associated with the movement is:

$$\Psi = \frac{1}{2} \ddot{x}_a(0)^2 - \dddot{x}_a(0) \cdot \dot{x}_a(0). \tag{3.20}$$

## 3.5 Non-Uniform Finger-Tapping

We now move on to more general, non-uniform finger-tapping models. In non-uniform finger-tapping, we allow that the distance travelled by the fingertip during each simple movement in the finger-tapping movement may vary from simple movement to simple movement, and that the time taken to perform each simple movement may vary from simple movement to simple movement. We do not try to provide any model for how the subject chooses distances and times for each simple movement. The aim is to construct a formalism using the optimal jerk-control model that we have developed in Chapter 1 that can describe observed finger-tapping performed by experimental subjects.

*3.5.1 Non-Uniform Basic Movement Finger-Tapping*

We can generalize the model for uniform basic movement finger-tapping to a model for non-uniform basic movement finger-tapping by replacing Eq. (3.7), where the extreme upward and downward positions and the timing is fixed, by a model where the fingertip moves through a sequence of extreme upward and downward positions $X_n$ at times $t_n$. We express this mathematically by finding the orbits $x_n(t)$ that minimize the cost functional and satisfy intermediate conditions given by:

$$\begin{aligned} J[x_n] &= \int_{T_n}^{T_{n+1}} L\left(\frac{1}{2}\ddot{x}_n^2 - \Phi(x_n, \dot{x}_n, \ddot{x}_n)\right) dt, \\ x_n(T_n) &= X_n, \quad x_n(T_{n+1}) = X_{n+1}, \\ \dot{x}_n(T_n) &= 0, \quad \dot{x}_n(T_{n+1}) = 0, \\ \ddot{x}_n(T_n) &= 0, \quad \ddot{x}_n(T_{n+1}) = 0. \end{aligned} \quad (3.21)$$

To keep the finger-tapping problem straightforward, we again assume the effect of $\Phi(x_n, \dot{x}_n, \ddot{x}_n)$ is small and take $\Phi(x_n, \dot{x}_n, \ddot{x}_n) = 0$. The Lagrangian for each finger-tapping movement becomes:

$$L = \frac{1}{2}\ddot{x}_n^2. \quad (3.22)$$

The Euler-Lagrange equation for Eq. (3.22) is again:

$$\dddot{x}_n = 0. \quad (3.23)$$

The generalized energy associated with each simple movement in the finger-tapping movement is:

$$\Psi_n = \frac{460^2 (X_{n+1} - X_n)^2}{2(T_{n+1} - T_n)^6}. \tag{3.24}$$

For an observed finger-tapping sequence consisting of $N$ simple movements, we can characterize the sequence using the average generalized energy using the formula:

$$\langle \Psi_n \rangle = \frac{460^2}{2N} \sum_{n=1}^{N} \frac{(X_{n+1} - X_n)^2}{(T_{n+1} - T_n)^6}. \tag{3.25}$$

An alternative is to characterize the sequence using the average of the square root of the generalized energy using the formula:

$$\langle \sqrt{\Psi_n} \rangle = \frac{460}{\sqrt{2}N} \sum_{n=1}^{N} \frac{|X_{n+1} - X_n|}{(T_{n+1} - T_n)^3}. \tag{3.26}$$

Eq. (3.26) provides a value close to the value of *amxfr* (amplitude × frequency) found to be useful in the analysis of finger-tapping in [43]. Modified slightly to reflect the model we have developed in this section, the value *amxfr* is:

$$amxfr = \frac{1}{N} \sum_{n=1}^{N} \frac{|X_{n+1} - X_n|}{T_{n+1} - T_n}. \tag{3.27}$$

We observe that Eq. (3.27) looks like an estimate for the average value for the square root of the classical energy of a simple harmonic oscillator (i.e. $\langle \sqrt{E} \rangle$ for $E$ in Eq. (3.6)).

*3.5.2 Non-Uniform Simple Movement Finger-Tapping*

We can generalize the model for non-uniform basic movement finger-tapping to a model for non-uniform simple movement finger-tapping by replacing Eq. (3.21), where we have assumed that the fingertip has zero acceleration at the extreme upward and downward positions, with a model in which accelerations $a_n$ are allowed at these times. We express this mathematically by finding the orbits $x_n(t)$ that minimize the cost functional and satisfy intermediate conditions given by:

$$J[x_n] = \int_{T_n}^{T_{n+1}} \left( \frac{1}{2} \ddot{x}_n^2 - \Phi(x_n, \dot{x}_n, \ddot{x}_n) \right) dt,$$
$$x_n(T_n) = X_n, \quad x_n(T_{n+1}) = X_{n+1},$$
$$\dot{x}_n(T_n) = 0, \quad \dot{x}_n(T_{n+1}) = 0, \tag{3.28}$$
$$\ddot{x}_n(T_n) = a_n, \quad \ddot{x}_n(T_{n+1}) = a_{n+1}.$$

As with the $X_n$ and $t_n$, we obtain the values $a_n$ by observation of the subject's actual finger-tapping. To keep the finger-tapping problem straightforward, we again assume the effect of $\Phi(x_n, \dot{x}_n, \ddot{x}_n)$ is small and take $\Phi(x_n, \dot{x}_n, \ddot{x}_n) = 0$. The Lagrangian for each finger-tapping movement becomes:

$$L = \frac{1}{2} \ddot{x}_n^2. \tag{3.29}$$

The Euler-Lagrange equation for Eq. (3.29) is again:

$$\ddddot{x}_n = 0. \tag{3.30}$$

We can solve for each simple movement; it will have the form:

$$x_n(t) = \sum_{v=0}^{5} c_v \cdot (t - T_n)^v. \tag{3.31}$$

We can evaluate the $c_v$ in Eq. (3.31) by requiring $x_n(t)$ to satisfy the intermediate conditions in Eq. (3.28), although we do not do that here. The generalized energy associated with each simple movement is:

$$\Psi_n = \frac{1}{2} \ddot{x}_n(T_n)^2 - \dddot{x}_n(T_n) \cdot \dot{x}_n(T_n) \tag{3.32}$$

For an observed finger-tapping sequence consisting of $N$ simple movements, we can characterize the sequence using the average generalized energy using the formula:

$$\langle \Psi_n \rangle = \frac{1}{N} \sum_{n=1}^{N} \Psi_n. \tag{3.33}$$

## 3.6 Discussion

We have presented an analysis of finger-tapping that treats the finger-tapping movement as a single complex movement made up of a sequence of simple movements using a number of models. We have described the simple movements making up finger-tapping using the optimal jerk-control model developed in Chapter 1. In Chapter 1, we found there is a quantity that is constant for the duration of a simple movement that minimizes the jerk-control cost functional – the generalized energy. In the analyses of finger-tapping that we have presented here, we characterized the performance of each simple movement in finger-tapping by the generalized energy associated with the simple movement. We then characterized the entire finger-tapping movement using the mean of the generalized energy of the component simple movements. These analyses bears a close relationship to the analysis of computer mouse movement in Chapter 2 where we have characterized how a subject performs a mouse movement in terms of how the subject chooses the generalized energy given the size of the target region being moved into. It also bears a close relationship to the analysis of finger-tapping in [43] that makes use of a quantity that is closely related to the generalized energy.

In Chapter 2, we characterized rapid, targeted movements of a computer mouse in terms of Fitts' law, which we approximated in the form:

$$T \approx \tau \cdot \left(\frac{D}{W}\right)^{1/3}. \tag{3.34}$$

We can look at Eq. (3.34) as characterizing how a subject moves a mouse in terms of how much total time $T$ is taken to make the mouse movement given the distance $D$ to the center of a target region of size $W$. We can rewrite Eq. (3.34) in the form:

$$\frac{W^2}{\tau^6} \approx \frac{D^2}{T^6}. \tag{3.35}$$

The right-hand side (RHS) of Eq. (3.35) bears a close relationship to the RHS of Eq. (3.13), which expresses the generalized energy of an individual simple movement within a uniform basic movement finger-tapping movement. Of course, there is nothing surprising about this given that we derived Eq. (3.34) by proposing that the generalized energy for a rapid, targeted movement has the form:

$$\Psi = \frac{1800}{\tau^6} W^2. \tag{3.36}$$

Thus, we find that by characterizing a subject's rapid, targeted mouse movements in terms of the parameter $\tau$, we have actually characterized the movements in terms of how the subject chooses the generalized energy of the movement given the size of the target. As a result, we can look at Fitts' law as a formula that characterizes rapid, targeted movements in terms of the generalized energy associated with the movement, and we find that Fitts' law characterizes the performance of rapid, targeted movements in more or less the same way that the generalized energy in Eqs. (3.13), (3.20), (3.25), and (3.33) characterizes the performance of finger-tapping.

As we noted, the value *amxfr* in Eq. (3.27) was used in [43] to characterizes finger-tapping performance in more detail than does the number of taps $N/2$ that the subject makes in 10s. It introduces the reasonable supposition of a relationship between the time intervals $T_{n+1} - T_n$ the fingertip takes to move, and the amplitudes $|X_{n+1} - X_n|$ moved by the fingertip, and that a metric used to characterize finger-tapping should account for this. In principle, any relationship between $T_{n+1} - T_n$ and $|X_{n+1} - X_n|$ that one proposes implies a physical model for finger-tapping. We have observed that we can interpret the choice of *amxfr* in Eq. (3.27) as being related to the classical energy of the motion where finger-tapping is described using the sinusoidal finger-tapping model (*amxfr* would actually related to mean of the square root of the classical energy). Instead of using the sinusoidal finger-tapping model, we have constructed finger-tapping as a complex movement built out of the simple movements described by the model in Chapter 1. In this model, a natural relationship between $T_{n+1} - T_n$ and $|X_{n+1} - X_n|$ appears in the form of the generalized energy.

In the models that we have developed to describe finger-tapping, we have treated finger-tapping as a complex movement that moves the fingertip between two extreme positions. However, in clinical tests of finger-tapping, the subject typically depresses a key or lever with each tap and the depression of the key or lever increments a counter. Thus, the in the downward portion of the finger-tapping movement the fingertip applies a force to an external, mechanical device and that mechanical device provides a hard limit to how far downward the fingertip may travel. Even if we assume that the force needed to depress the key or lever is negligible, there is still the issue of the hard limit on the downward movement. This means the subject does not need to make a movement that slows to a stop when moving downward, but can rely on the device and table to bring the downward movement to a stop. A more detailed model of finger-tapping would look at how the subject might modify the finger-tapping movement given the existence of this limit to achieve a higher finger-tapping speed. We would expect that such an analysis would clarify the relationship between finger-tapping and typing on a keyboard.

We have shown that, in principle, the subject is free to choose the value for the acceleration of the finger-tip at the instant when one simple movement ends and the next begins, while still having the finger-tapping take on a basically oscillatory form. We have presented models in which the subject simply sets the acceleration to zero at these times (basic movement finger-tapping) or finds the value that minimizes the overall cost of the complex finger-tapping movement (minimum cost simple movement finger-tapping). The actual value of the acceleration at these times needs to be determined empirically and likely depends on a more detailed analysis of the finger-tapping experiment including how the subject moves when there is a lever or button to be depressed and a hard stop at the downward limit of the finger-tapping movement.

# Chapter 4 – Tremor-Like Solutions to the Optimal Jerk-Control Model of Human Movements

**Abstract.** In Chapter 1, we have developed optimal jerk-control model to describe simple movements. In this Chapter, we show that, when we define the cost functional for this model of simple movements using parameter values in a particular range, the movement that minimizes the cost functional exhibits an oscillatory motion that resembles tremor. We examine how the frequency and amplitude of the oscillatory motion in these tremor-like solutions to the optimal jerk-control problem arise out of the parameter values defining the cost functional and the initial and final conditions that the movement must satisfy. We also look at the form of the underlying movement when we remove the effect of the oscillatory motion from the tremor-like solution and note that it takes on the approximate form of a simple movement determined by minimizing the cost functional of an acceleration-control model. We note that the resemblance of the underlying movement to one arising from an acceleration-control model means that this underlying movement would appear to the eye to be less smooth than a movement arising from the jerk-control model. Finally, we look at how we can relate the tremor-like solutions to the optimal jerk-control model to two broad classes of tremor − (1) action tremor, and (2) resting tremor.

## 4.1 Introduction

In Chapters 1, 2, and 3, we have developed an optimal jerk-control model and used to describe simple movements, that is, movements that minimize a cost functional while satisfying a set of initial and final conditions. In all three Chapters, the mathematical description of the simple movements took the form of a smooth movement going from the initial to the final conditions. While a model that describes movements using smooth movements is useful for a wide range of healthy human movements, there is a range of human movements that are not smooth, including movements exhibiting a tremor – an involuntary oscillatory movement of a body part. In this Chapter, we show that the optimal jerk-control model for describing human movement that we have developed in Chapter 1 is already capable, with no further modification, of describing simple movements that exhibit an oscillatory character that resembles tremor. These solutions arise from cost functionals with parameter values lying in a particular range.

**4.2 Tremor-Like Solutions of the Optimal Jerk-Control Model**

We begin by showing how tremor-like solutions arise from the optimal jerk-control model. We recall that, in the model that we have developed in Chapter 1, the subject controls the motion of a specific point on the body (e.g. a point on the hand). In the model, it is assumed that the body is constrained in such a way that the motion of this point determines the motion of the entire body.

In Chapter 1, we formulated the optimal jerk-control model of movement as that of finding the orbit $x(t)$ of the point on the body being controlled that minimizes the cost functional $J[\cdot]$ subject to the initial and final conditions specified by:

$$
\begin{aligned}
J[x] = \int_0^T &\left( \frac{1}{2}\dddot{x}^2 - \Phi(x, \dot{x}, \ddot{x}) \right) dt, \\
x(0) &= -D, \quad x(T) = 0, \\
\dot{x}(0) &= v_0, \quad \dot{x}(T) = v_1, \\
\ddot{x}(0) &= a_0, \quad \ddot{x}(T) = a_1.
\end{aligned}
\tag{4.1}
$$

We restrict our analysis to basic movements, that is, simple movements that begin and end at rest. We replace the initial and final conditions in Eq. (4.1) with the initial and final conditions:

$$\begin{aligned} x(0) &= -D, & x(T) &= 0, \\ \dot{x}(0) &= 0, & \dot{x}(T) &= 0, \\ \ddot{x}(0) &= 0, & \ddot{x}(T) &= 0. \end{aligned} \quad (4.2)$$

We assume that we can approximate $\Phi(x, \dot{x}, \ddot{x})$ using a Taylor series expansion truncated to second-order. Following the Chapter 1, it suffices to use the *effective Lagrangian* for deriving the orbit $x(t)$ that minimizes the cost functional in Eq. (4.1):

$$L_{eff} = \frac{1}{2}\ddot{x}^2 - \phi_{01} x - \frac{1}{2} \begin{bmatrix} x \\ \dot{x} \\ \ddot{x} \end{bmatrix}^T \begin{bmatrix} \phi_{11} & 0 & \phi_{13} \\ 0 & \phi_{22} & 0 \\ 0 & 0 & \phi_{33} \end{bmatrix} \begin{bmatrix} x \\ \dot{x} \\ \ddot{x} \end{bmatrix}. \quad (4.3)$$

The parameters $\phi_{01}$, $\phi_{11}$, and $\phi_{13}$ describe a cost associated with the position of the point being moved, and, in Chapter 2, we used these parameters to describe the largely cognitive process of make a movement that ends in a specified target region. To keep the analysis straightforward, we assume that the effect of the parameters $\phi_{01}$, $\phi_{11}$, and $\phi_{13}$ is negligible (i.e. there is no specified target region being move into, or the target region is very large). In this case, the effective Lagrangian becomes:

$$L_{eff} \approx \frac{1}{2}\ddot{x}^2 - \frac{1}{2} \begin{bmatrix} x \\ \dot{x} \\ \ddot{x} \end{bmatrix}^T \begin{bmatrix} 0 & 0 & 0 \\ 0 & \phi_{22} & 0 \\ 0 & 0 & \phi_{33} \end{bmatrix} \begin{bmatrix} x \\ \dot{x} \\ \ddot{x} \end{bmatrix}. \quad (4.4)$$

The subject moves the point on the body that is being controlled using the orbit $x(t)$ that minimizes the cost functional in Eq. (4.1) subject to the initial and final conditions in Eq. (4.2) where the integrand is Eq. (4.4). This orbit is the one that satisfies the Euler-Lagrange equation for Eq. (4.4), that is:

$$\dddot{x} + \phi_{33}\ddot{x} - \phi_{22}\ddot{x} = 0. \tag{4.5}$$

We now investigate the behavior of solutions to Eq. (4.5) for particular ranges of values for the parameters $\phi_{22}$ and $\phi_{33}$. We show that for choices of parameter values $\phi_{22} > 0$ and $\phi_{33} > 0$, the solutions to Eq. (4.5) take on an oscillatory and tremor-like character.

*4.2.1 Case 1: $\phi_{22} = \phi_{33} = 0$*

We begin by looking at the case where $\phi_{22} = \phi_{33} = 0$. The solutions to Eq. (4.5) do not take on oscillatory or tremor-like character in this case. We investigate this case to look at the basic form of the optimal jerk-control movement so that we can see how non-zero values of the parameters $\phi_{22}$ and $\phi_{33}$ change the orbit $x(t)$ of the movement. In this case, Eq. (4.5) becomes:

$$\dddot{x} = 0. \tag{4.6}$$

Integrating (4.6) six times, we find the optimal orbit $x(t)$ of the movement to have the form:

$$x(t) = c_0 + c_1 t + c_2 t^2 + c_3 t^3 + c_4 t^4 + c_5 t^5. \tag{4.7}$$

We calculate the constants $c_n$ by requiring the movement to satisfy the initial and final conditions given in Eq. (4.2).

We note that we have used models with the Euler-Lagrange equation in Eq. (4.6) in Chapter 2 to arrive at the approximate expression $T = \tau \cdot (D/W)^{1/3}$ for Fitts' law, and in Chapter 3 to describe finger-tapping.

*4.2.2 Case 2: $\phi_{22} = 0$*

For the case where $\phi_{22} = 0$, Eq. (4.5) becomes:

$$\dddot{x} + \phi_{33}\ddot{x} = 0. \tag{4.8}$$

Integrating Eq. (4.8) four times, we find the optimal orbit $x(t)$ of the movement to have the form:

$$\ddot{x} + \phi_{33} x = k_0 + k_1 t + k_2 t^2 + k_3 t^3. \tag{4.9}$$

Eq. (4.9) is a second-order non-homogeneous differential equation. We begin the process of solving the non-homogeneous differential equation by first solving the corresponding homogeneous differential equation. The second-order homogeneous differential equation corresponding to Eq. (4.9) is:

$$\ddot{x} + \phi_{33} x = 0. \tag{4.10}$$

We can think of the right-hand side (RHS) of the non-homogeneous differential equation in Eq. (4.9) as driving the system described by the left-hand side (LHS); solving the homogeneous differential equation in Eq. (4.10) provides information about how the system behaves in general. We define the angular frequency $\omega = \sqrt{|\phi_{33}|}$. For purposes of illustration and comparison, we choose an arbitrary value of $\omega = 6.6845\pi$ for the figures.

In the case $\phi_{33} > 0$, the solution to the homogeneous differential equation in Eq. (4.10) is:

$$x(t) = c_4 \cos(\omega t) + c_5 \sin(\omega t). \tag{4.11}$$

The corresponding solution to the non-homogeneous differential equation in Eq. (4.9) has the form:

$$x(t) = c_0 + c_1 t + c_2 t^2 + c_3 t^3 + c_4 \cos(\omega t) + c_5 \sin(\omega t). \tag{4.12}$$

We calculate the constants $c_n$ by requiring the movement to satisfy the initial and final conditions given in Eq. (4.2). The presence of $\cos(\omega t)$ and $\sin(\omega t)$ in Eq. (4.12) gives the orbit $x(t)$ an oscillatory or tremor-like character. We can decompose Eq. (4.12) into the sum of a term $x_M(t)$ expressing the movement from the initial to the final positions and a term $x_T(t)$ describing the tremor-like component of the movement:

$$\begin{aligned} x(t) &= x_M(t) + x_T(t), \\ x_M(t) &= c_0 + c_1 t + c_2 t^2 + c_3 t^3, \\ x_T(t) &= c_4 \cos(\omega t) + c_5 \sin(\omega t). \end{aligned} \quad (4.13)$$

The $x_M(t)$ term gives the approximate form of orbit if we somehow were to filter out the effect of the tremor-like component. The orbit of the movement $x_M(t)$ in Eq. (4.13) looks like the optimal orbit for the case of acceleration-control where the effective Lagrangian is:

$$L_{eff} = \frac{1}{2} \ddot{x}_M{}^2. \quad (4.14)$$

The motion of the tremor-like movement looks like the motion of a simple harmonic oscillator whose point of equilibrium at any time $t$ is $x_M(t)$. The orbit of the tremor-like movement $x_T(t)$ in Eq. (4.13) looks like an optimal orbit for the case of velocity-control where the effective Lagrangian is:

$$L_{eff} = \frac{1}{2} \dot{x}_T{}^2 - \frac{1}{2} \omega^2 x_T{}^2. \quad (4.15)$$

The orbit $x(t)$ looks something like a movement made using acceleration-control that has oscillatory motion present within it. We observe that Eq. (4.15) suffices to determine the angular frequency $\omega$ of the tremor, but not the amplitude. The amplitude is determined by the initial and final conditions in the original jerk-control movement problem. We provide a comparison of the functional form of the model in Eq. (4.12) to Eq. (4.7) for this case in Figure 4.1.

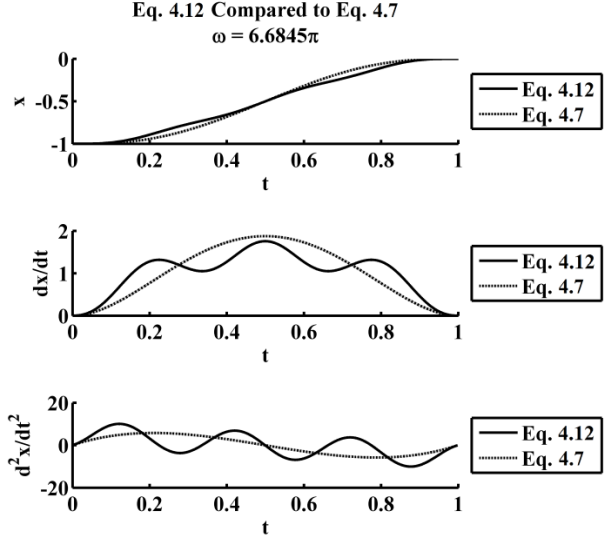

**Figure 4.1.** Eq. (4.12) compared to Eq. (4.7). The model in Eq. (4.12) is evaluated for $\omega = 6.6845\pi$. The orbit ($x$), velocity of the orbit ($dx/dt$), and acceleration of the orbit ($dx^2/dt^2$) are shown in separate plots. The movement begins at time $t = 0$ and end at time $t = 1$.

The acceleration-control movement that we would observe were we to somehow filter out the effect of the tremor would appear less smooth than the corresponding jerk-control movement described in Eq. (4.7). As we have observed in Chapter 1, in the general control model of which acceleration- and jerk-control are special cases, the control can be set to an arbitrary value at the beginning and end of the movement. Thus, the acceleration-control movement would exhibit a rather large and sudden change in acceleration at the beginning and end of the movement, rather than the more smooth change that a jerk-control movement would exhibit. This would correspond to a large and sudden change in the force exerted by the muscles at the beginning and end of the movement.

In the case $\phi_{33} < 0$, the solution to the homogeneous differential equation in Eq. (4.10) is:

$$x = c_4 \cosh(\omega t) + c_5 \sinh(\omega t). \tag{4.16}$$

The corresponding solution to the non-homogeneous differential equation in Eq. (4.9) has the form:

$$x(t) = c_0 + c_1 t + c_2 t^2 + c_3 t^3 + c_4 \cosh(\omega t) + c_5 \sinh(\omega t). \tag{4.17}$$

We calculate the constants $c_n$ by requiring the movement to satisfy the initial and final conditions given in Eq. (4.2). The orbit $x(t)$ does not exhibit an oscillatory or tremor-like behavior. We provide a comparison of the functional form of the model in Eq. (4.17) to Eq. (4.7) for this case in Figure 4.2.

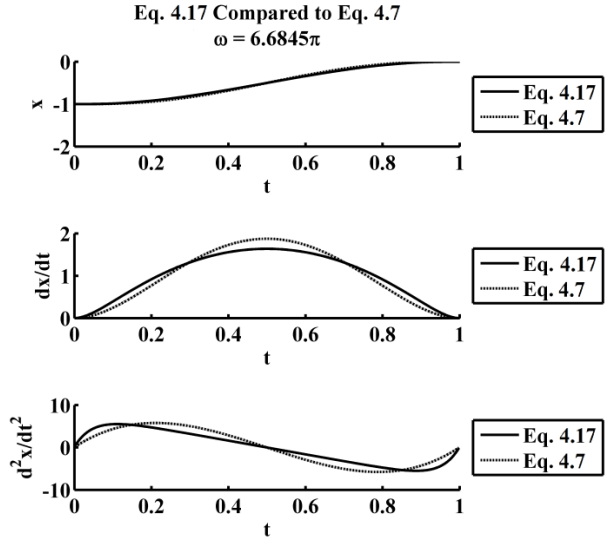

**Figure 4.2.** Eq. (4.17) compared to Eq. (4.7). The model in Eq. (4.17) is evaluated for $\omega = 6.6845\pi$. The orbit ($x$), velocity of the orbit ($dx/dt$), and acceleration of the orbit ($dx^2/dt^2$) are shown in separate plots. The movement begins at time $t = 0$ and end at time $t = 1$.

### 4.2.3 Case 3: $\phi_{33} = 0$

For the case where $\phi_{33} = 0$, Eq. (4.5) becomes:

$$\dddot{x} - \phi_{22}\ddot{x} = 0. \tag{4.18}$$

Integrating Eq. (4.18) twice, we find the optimal orbit $x(t)$ of the movement to have the form:

$$\ddddot{x} - \phi_{22}x = k_0 + k_1 t. \tag{4.19}$$

Eq. (4.19) is a fourth-order non-homogeneous differential equation. We begin the process of solving the non-homogeneous differential equation by first solving the corresponding homogeneous differential equation. The fourth-order homogeneous differential equation corresponding to Eq. (4.19) is:

$$\ddddot{x} - \phi_{22}x = 0. \tag{4.20}$$

We define the angular frequency $\omega = \sqrt[4]{|\phi_{22}|}$. For purposes of illustration and comparison, we choose an arbitrary value of $\omega = 6.6845\pi$ for the figures.

In the case $\phi_{22} > 0$, the solution to the homogeneous differential equation in Eq. (4.20) has the form:

$$x(t) = c_2 \cosh(\omega t) + c_3 \sinh(\omega t) + c_4 \cos(\omega t) + c_5 \sin(\omega t). \tag{4.21}$$

The corresponding solution to the non-homogeneous differential equation in Eq. (4.19) has the form:

$$x(t) = c_0 + c_1 t + c_2 \cosh(\omega t) + c_3 \sinh(\omega t) + c_4 \cos(\omega t) + c_5 \sin(\omega t). \tag{4.22}$$

We calculate the constants $c_n$ by requiring the movement to satisfy the initial and final conditions given in Eq. (4.2). The presence of $\cos(\omega t)$ and $\sin(\omega t)$ in Eq. (4.22) gives the orbit $x(t)$ an oscillatory or tremor-like character. We can decompose Eq. (4.22) into the sum of a term $x_M(t)$ expressing the movement from the initial to the final positions and a term $x_T(t)$ describing the tremor in the movement:

$$x(t) = x_M(t) + x_T(t),$$
$$x_M(t) = c_0 + c_1 t + c_2 \cosh(\omega t) + c_3 \sinh(\omega t), \qquad (4.23)$$
$$x_T(t) = c_4 \cos(\omega t) + c_5 \sin(\omega t).$$

The $x_M(t)$ term gives the approximate form of orbit if we somehow were to filter out the effect of the tremor. The orbit of the movement $x_M(t)$ in Eq. (4.23) looks like the optimal orbit for the case of acceleration-control where the effective Lagrangian is:

$$L_{eff} = \frac{1}{2}\ddot{x}_M^2 + \frac{1}{2}\omega^2 \dot{x}_M^2. \qquad (4.24)$$

The motion of the tremor-like movement looks like the motion of a simple harmonic oscillator whose point of equilibrium at any time $t$ is $x_M(t)$. The orbit of the tremor-like movement $x_T(t)$ in Eq. (4.23) looks like an optimal orbit for the case of velocity-control where the effective Lagrangian is:

$$L_{eff} = \frac{1}{2}\dot{x}_T^2 - \frac{1}{2}\omega^2 x_T^2. \qquad (4.25)$$

The orbit $x(t)$ looks something like a movement made using acceleration-control that has oscillatory motion present within it. We again observe that Eq. (4.25) suffices to determine the angular frequency $\omega$ of the tremor, but not the amplitude. We provide a comparison of the functional form of the model in Eq. (4.22) to Eq. (4.7) for this case in Figure 4.3.

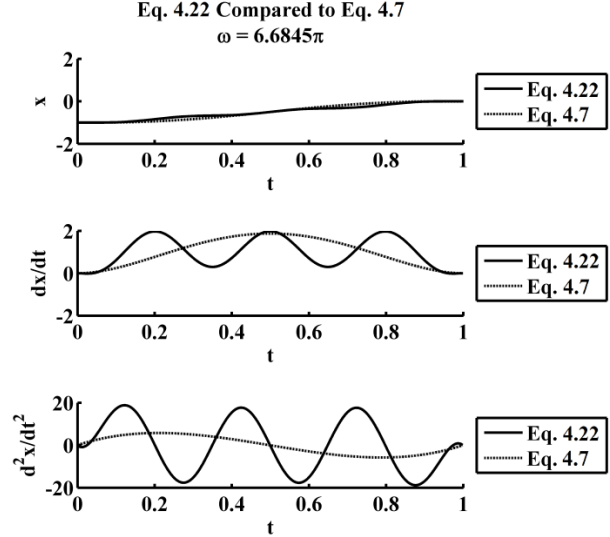

**Figure 4.3.** Eq. (4.22) compared to Eq. (4.7). The model in Eq. (4.22) is evaluated for $\omega = 6.6845\pi$. The orbit ($x$), velocity of the orbit ($dx/dt$), and acceleration of the orbit ($dx^2/dt^2$) are shown in separate plots. The movement begins at time $t = 0$ and end at time $t = 1$.

The acceleration-control movement that we would observe were we somehow to filter out the effect of the tremor would again appear less smooth than the corresponding jerk-control movement described in Eq. (4.7). It would again exhibit a rather large and sudden change in acceleration at the beginning and end of the movement, rather than the more smooth change that a jerk-control movement would exhibit.

In the case $\phi_{22} < 0$, the solution to the homogeneous differential equation in Eq. (4.20) has the form:

$$x(t) = c_2 \cosh\left(\frac{\sqrt{2}}{2}\omega t\right)\cos\left(\frac{\sqrt{2}}{2}\omega t\right) + c_3 \cosh\left(\frac{\sqrt{2}}{2}\omega t\right)\sin\left(\frac{\sqrt{2}}{2}\omega t\right)$$
$$+ c_4 \sinh\left(\frac{\sqrt{2}}{2}\omega t\right)\cos\left(\frac{\sqrt{2}}{2}\omega t\right) + c_5 \sinh\left(\frac{\sqrt{2}}{2}\omega t\right)\sin\left(\frac{\sqrt{2}}{2}\omega t\right).$$

(4.26)

The corresponding solution to the non-homogeneous differential equation in Eq. (4.19) has the form:

$$x(t) = c_0 + c_1 t$$
$$+ c_2 \cosh\left(\frac{\sqrt{2}}{2}\omega t\right)\cos\left(\frac{\sqrt{2}}{2}\omega t\right) + c_3 \cosh\left(\frac{\sqrt{2}}{2}\omega t\right)\sin\left(\frac{\sqrt{2}}{2}\omega t\right) \quad (4.27)$$
$$+ c_4 \sinh\left(\frac{\sqrt{2}}{2}\omega t\right)\cos\left(\frac{\sqrt{2}}{2}\omega t\right) + c_5 \sinh\left(\frac{\sqrt{2}}{2}\omega t\right)\sin\left(\frac{\sqrt{2}}{2}\omega t\right).$$

We calculate the constants $c_n$ by requiring the movement to satisfy the initial and final conditions given in Eq. (4.2). Despite the presence of $\cos(\omega t)$ and $\sin(\omega t)$ in Eq. (4.27), the orbit $x(t)$ does not exhibit an oscillatory or tremor-like behavior. We provide a comparison of the functional form of the model in Eq. (4.27) to Eq. (4.7) for this case in Figure 4.4.

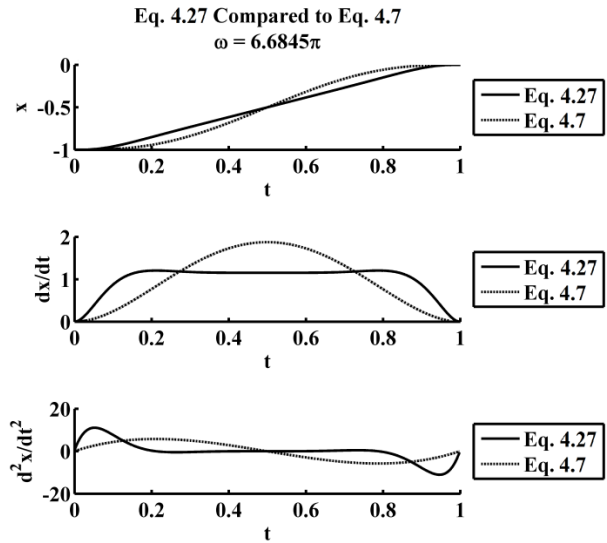

**Figure 4.4.** Eq. (4.27) compared to Eq. (4.7). The model in Eq. (4.27) is evaluated for $\omega = 6.6845\pi$. The orbit ($x$), velocity of the orbit ($dx/dt$), and acceleration of the orbit ($dx^2/dt^2$) are shown in separate plots. The movement begins at time $t = 0$ and end at time $t = 1$.

**4.3 Tremor**

We have shown how for parameter values $\phi_{22} > 0$ and $\phi_{33} > 0$, the orbit $x(t)$ that minimizes the jerk-control cost functional $J[\cdot]$ exhibits an oscillatory or tremor-like behavior. We now look at two broad classes of tremor – (1) action tremor, and (2) resting tremor – and examine how we might describe tremors of these classes in terms of tremor-like orbits in the optimal jerk-control model.

*4.3.1 Action Tremor*

Action tremors occur during voluntary movement of a body part. Mathematically, we treat action tremors by looking at a sequence of three movements: (1) the point on the body being controlled being at rest, (2) the subject moving the point from one position to another, and (3) the point being at rest. We can think of a point being at rest for some period as an optimal movement in which the beginning and ending positions are the same and the movement begins and ends at rest. The optimal orbit $x(t)$ that satisfies this problem is one in which the point being controlled does not move. We require that there be no tremor-like behavior during the movements first and third movements (when the point being controlled is at rest), but that a tremor-like behavior appears during the second movement (when the point being controlled is moving).

For movements in one dimension, we have solved cases of this model of action tremor in Sections 4.2.2 and 4.2.3. In these cases, we have treated the movement as a basic movement, so it begins and ends at rest. We can make extend this to the model of action tremor by requiring that the periods of rest at the beginning and end of the movement extend over some time interval.

*4.3.2 Resting Tremor*

Resting tremor occurs when the body part is at rest. We look at the asymmetric case where the tremor is present during rest but becomes less prominent during voluntary movement. To keep the mathematics straightforward, we treat the tremor as vanishing during voluntary movement. Mathematically, we treat tremor tremors by looking at a sequence of three movements: (1) the subject moving the point on the body being controlled from one position to another, (2) the point being at rest, and (3) the subject moving the point from one position to another. Thus, we treat resting tremor as the inverse of action tremor. We again can think of a point being at rest for some period as an optimal movement in which the beginning and ending positions are the same and the movement begins and ends at rest. We require that there be no tremor-like behavior during the movements first and third movements (when the point being controlled is moving), but that a tremor-like behavior appears during the second movement (when the point being controlled is at rest).

We can extrapolate the model of tremor in Sections 4.2.2 and 4.2.3 to the case of resting tremor by treating the orbit of the movement $x_M(t)$ as being the optimal acceleration control movement that begins and ends at rest. Thus, the motion of the resting tremor is simply described by the orbit of the simple harmonic oscillator $x_T(t)$.

*4.3.3 Remarks*

In Chapter 1, we noted that the jerk can have, and typically does have, a discontinuity at the instant in time when a simple movement begins or ends. We observe that, in the model of action tremor, the tremor begins at the discontinuity in the jerk at the beginning of a voluntary movement, and ends at the discontinuity in the jerk at the end of the voluntary movement. Similarly, we observe that, in the model of resting tremor, the tremor begins at the discontinuity in the jerk at the end of a voluntary movement, and begins at the discontinuity in the jerk at the beginning of the next voluntary movement. A discontinuity in the jerk corresponds to a cusp in the functional form of the acceleration of the orbit $\ddot{x}(t)$, where the derivative of the acceleration is not defined, and thus it corresponds to a cusp in the functional form of the net force applied by the muscles where the derivative of the net force is not defined. Thus, we find that these cusps correspond to instants where tremors are triggered or where they are stopped.

We can construct \models of tremor containing both action and resting tremor. As in the case of pure action and resting tremor (in the sense we have given them), there would be a transition between action and resting tremor where cusps occur in the acceleration of the orbit. We can also extend our treatment of tremor, which we have developed for basic movements, to complex movements consisting of a sequence of simple movements. Complex movements can involve a sequence of cusps within the acceleration of the orbit corresponding to transitions from one simple movement to another; we may find a progression in the nature of the tremor through the sequence of cusps of the complex movement.

**4.4 Discussion**

We have shown that the model of simple human movements that we have developed in Chapter 1 is capable of producing tremor-like solutions with no further modification. The tremor-like solutions simply arise due to particular values of the parameters present in the function $\Phi(x,\dot{x},\ddot{x})$ in the cost functional $J[\cdot]$. We have found, that were the oscillations in the tremor-like solutions to be filtered out, the remaining movement resembles one made using an acceleration-control model, and would appear to be less smooth than corresponding movement made using a jerk-control model. We have also shown that we can model to basic types of tremor – (1) action tremor, and (2) resting tremor – by treating the tremor as being triggered and stopped by a cusp in the acceleration $\ddot{x}(t)$; that is, being triggered and stopped at instants in time when the jerk is discontinuous.

We have observed that the angular frequency $\omega$ of the tremor-like solution is determined by the parameter values in the function $\Phi(x,\dot{x},\ddot{x})$, while the amplitude of the tremor-like solution is determined by this together with the initial and final conditions in the original jerk-control movement problem. Thus, tremor-like solutions have different amplitudes of oscillation depending on the movement that the subject is carrying out. Moreover, in the case where at tremor-like solution exists both before and after cusp in the acceleration (i.e. discontinuity in the jerk) we may find the amplitude of the tremor is different on either side of the cusp. We can think of action and resting tremors as we have modeled them as extreme cases of this where the amplitude of the oscillation goes to zero on one side of the cusp. We have refrained from working out, in this Chapter, a general treatment of how the amplitude of the oscillation in the tremor-like solution is related to these other conditions.